\documentclass[reprint, superscriptaddress, amsmath,amssymb, aps, prb]{revtex4-2}
\usepackage{graphicx}
\usepackage{dcolumn}
\usepackage{bm}
\usepackage{hyperref}
\usepackage[utf8x]{inputenc}
\usepackage{upgreek}
\usepackage{color, soul}

\begin{document}

\title{Bloch-to-N\'eel domain wall transition evinced through morphology of magnetic bubble expansion in Ta/CoFeB/MgO layers}

\author{Arianna Casiraghi}
\email{a.casiraghi@inrim.it}
\affiliation{Istituto Nazionale di Ricerca Metrologica, Strada delle Cacce 91, 10135 Torino, Italy}

\author{Alessandro Magni}
\affiliation{Istituto Nazionale di Ricerca Metrologica, Strada delle Cacce 91, 10135 Torino, Italy}

\author{Liza Herrera Diez}
\affiliation{Centre de Nanosciences et de Nanotechnologies, CNRS, Univ. Paris-Sud, Universit\'e Paris-Saclay, C2N Orsay, 91405 Orsay cedex, France}

\author{Juergen Langer}
\affiliation{Singulus Technology AG, Hanauer Landstrasse 103, 63796 Kahl am Main, Germany}

\author{Berthold Ocker}
\affiliation{Singulus Technology AG, Hanauer Landstrasse 103, 63796 Kahl am Main, Germany}

\author{Massimo Pasquale}
\affiliation{Istituto Nazionale di Ricerca Metrologica, Strada delle Cacce 91, 10135 Torino, Italy}

\author{Dafin\'e Ravelosona}
\affiliation{Centre de Nanosciences et de Nanotechnologies, CNRS, Univ. Paris-Sud, Universit\'e Paris-Saclay, C2N Orsay, 91405 Orsay cedex, France}

\author{Gianfranco Durin}
\affiliation{Istituto Nazionale di Ricerca Metrologica, Strada delle Cacce 91, 10135 Torino, Italy}

\date{\today}

\begin{abstract}
Ta/CoFeB/MgO trilayers with perpendicular magnetic anisotropy are often characterised by vanishing or modest values of interfacial Dzyaloshinskii-Moriya interaction (DMI), which results in purely Bloch or mixed Bloch-N\'eel domain walls (DWs). Here we investigate the creep evolution of the overall magnetic bubble morphology in these systems under the combined presence of in-plane and out-of-plane magnetic fields
and we show that He$^+$ ion irradiation induces a transition of the internal DW structure towards a fully N\'eel spin texture. This transition can be correlated to a simultaneous increase in DMI strength and reduction in saturation magnetisation -- which are a direct consequence of the effects of ion irradiation on the bottom and top CoFeB interfaces, respectively. The threshold irradiation dose above which DWs acquire a pure N\'eel character is experimentally found to be between 12 $\times$ 10$^{18}$ He$^+$/m$^2$ and 16 $\times$ 10$^{18}$ He$^+$/m$^2$, matching estimations from the one dimensional DW model based on material parameters. Our results indicate that evaluating the global bubble shape during its expansion can be an effective tool to sense the internal bubble DW structure. Furthermore, we show that ion irradiation can be used to achieve post-growth engineering of a desired DW spin texture.
\end{abstract}

                             
\maketitle

\section{\label{sec:introduction} Introduction}
In magnetic systems with spin-orbit coupling and lack of structural inversion symmetry, the Dzyaloshinskii-Moriya interaction \cite{Dzyaloshinsky_JPhysChemSolids_1958, Moriya_PR_1960} (DMI) plays a crucial role in stabilising topological spin textures, most notably magnetic skyrmions and chiral domain walls (DWs). While these spin structures were initially discovered in chiral magnetic single crystals \cite{Rossler_Nature_2006, Muhlbauer_Science_2009, Yu_Nature_2010}, research on the topic has more recently been extended to ferromagnetic ultra-thin films and multilayers \cite{Heinze_NatPhys_2011, Emori_NatMater_2013, Ryu_NatNanotechnol_2013, Moreau-Luchaire_NatNanotechnol_2016}, which are more suitable for technological applications. In this context, systems investigated are composed of a heavy metal (HM) layer and an adjacent ultra-thin ferromagnetic (FM) film with perpendicular magnetic anisotropy. Here the DMI has an interfacial nature \cite{Bode_Nature_2007}, since it arises due to the broken inversion symmetry at the interface between the two materials, while the spin-orbit coupling originates in the HM atoms, which mediate the interaction between neighbouring FM spins. Acting like an effective in-plane magnetic field, the interfacial DMI promotes the stabilisation of N\'eel DWs with a fixed chirality \cite{Thiaville_EPL_2012}, while achiral Bloch DWs would otherwise be favoured in these ultra-thin films from purely magnetostatic considerations. Recent experiments have demonstrated that N\'eel DWs can be driven by spin-orbit torques with an unprecedented level of efficiency \cite{Emori_NatMater_2013, Ryu_NatNanotechnol_2013, Yang_NatNanotechnol_2015}, making them promising candidates as information carriers in storage and logic devices. This has motivated extensive research in trying to quantify the strength of the interfacial DMI \cite{Je_PRB_2013, Torrejon_NatCommun_2014, Franken_SciRep_2014, Benitez_NatCommun_2015, Nembach_NatPhys_2015, Woo_NatMater_2016}, which is directly linked to the internal DW structure.

The strength and sign of the DMI are primarily dictated by the stack composition, which typically consists in the FM film being sandwiched either between two HM layers (e.g. Ta, W, Pt, Ir, $\ldots$), or between a HM layer and a non metallic one (e.g. MgO, AlOx, $\ldots$). Furthermore, deposition conditions and parameters are also critical aspects in defining DMI strength and sign, since they can have a profound impact on the interface roughness and/or intermixing \cite{Hrabec_PRB_2014, Lavrijsen_PRB_2015, Wells_PRB_2017}. In any case, once the stack has been grown, its DMI is fixed and consequently also DWs have a precise internal structure. Indeed, in order to tune the DW chirality from a mixed Bloch-N\'eel state into a fully N\'eel texture, some works have resorted to altering the original stack composition by inserting thin material layers of varying thicknesses between the FM layer and either the top or the bottom layer \cite{Hrabec_PRB_2014, Chen_NatCommun_2013, Chen_AdvMater_2015, Cao_Nanoscale_2018}. A more practical approach to tailor DW chirality would require some form of \textit{post-growth} treatment. So far, only a few studies have explored post-growth strategies to tune the DMI strength: these have involved either annealing the stack at different temperatures \cite{Khan_APL_2016}, or using ion irradiation to engineer interface intermixing \cite{Balk_PRL_2017, HerreraDiez_PRB_2019}. Yet, while these studies clearly indicate that post-growth tuning of DMI is attainable, they do not address the correlation between DMI strength and internal DW structure, which is particularly important in the transition from a Bloch to a N\'eel DW.  

In this work we investigate the influence of He$^+$ ion irradiation on Ta/CoFeB/MgO trilayers and we show that, by finely tuning interface disorder via irradiation, it is possible to induce a crossover from DWs with mostly Bloch character to purely N\'eel ones. In particular, we explore the creep dynamics of magnetic bubble domains under simultaneous application of in-plane and out-of-plane magnetic fields, which is one of the most conventional techniques used to quantify the interfacial DMI in these or similar stacks \cite{Je_PRB_2013, Hrabec_PRB_2014, Khan_APL_2016}. But rather than inspecting DW velocities along the direction of the in-plane field only, as it is most commonly done, we analyse how the overall morphology of the bubble domains changes during expansion. In this way we are able to clearly observe the tuning of DW chirality towards a fully N\'eel texture induced by an increasing irradiation dose, which in turns corresponds to both an increased intermixing at the Ta/CoFeB interface \cite{HerreraDiez_PRB_2019} and a decreased Fe content at the CoFeB/MgO interface \cite{HerreraDiez_APL_2015}. The threshold irradiation dose for which the bubble shape evolution corresponds to DWs in the N\'eel state is found to be between 12 $\times$ 10$^{18}$ He$^+$/m$^2$ and 16 $\times$ 10$^{18}$ He$^+$/m$^2$, which matches with predictions from the one dimensional DW model \cite{Thiaville_EPL_2012} based on the measured materials parameters. Our results indicate the importance of evaluating the overall bubble domain morphology to extract information on the internal DW texture. Furthermore, we demonstrate that interface engineering can be used to tailor the DW structure via post-growth ion irradiation, opening new routes for designing spintronics devices.

\section{\label{sec:methods} Methods}
The samples investigated consist of Ta(5nm)/Co$_{20}$Fe$_{60}$B$_{20}$(1nm)/MgO(2nm)/Ta(3nm) films, which were grown onto thermally oxidised silicon substrates by magnetron sputtering. The samples were subsequently annealed at 300 $^{\circ}$C and exposed to irradiation doses (ID) ranging from 4 $\times$ 10$^{18}$ He$^+$/m$^2$ up to 16 $\times$ 10$^{18}$ He$^+$/m$^2$. The ion irradiation process was performed at room temperature and at a constant energy of 15 keV. The sketch in Fig.~\ref{Fig1}(a) illustrates the sample structure and the effect of the ion irradiation process on the interfaces. All samples exhibit perpendicular magnetic anisotropy (PMA), with saturation magnetisation ($M_\mathrm{S}$) and effective perpendicular anisotropy constant ($K_\mathrm{eff}$) decreasing approximately linearly with increasing ID, as reported in a previous work \cite{HerreraDiez_APL_2015}. The exchange stiffness constant ($A$) for the pristine sample was estimated from that of pure Fe and Co samples, as well as from values reported in the literature \cite{Yamanouchi_IEEEMagnLett_2011}. As previously reported \cite{HerreraDiez_PRB_2019}, $A$ for the irradiated samples was evaluated assuming a weak dependence on $M_\mathrm{S}^2$ , in agreement with predictions and experimental observations \cite{Asti_PRB_2007, Hubert}. The values of $M_\mathrm{S}$, $K_\mathrm{eff}$, and $A$ for each of the five samples investigated are listed in Tab.~\ref{Tab1}. 

Magnetic bubble domain expansion in the creep regime was imaged using a wide-field magneto-optical Kerr effect (MOKE) microscope in polar configuration, equipped with an in-plane (IP) electromagnet and a coil for out-of-plane (OOP) field pulses. A bubble domain was initially nucleated in the pre-saturated sample by applying a short OOP field pulse. The bubble domain was later expanded either with an OOP pulse only, or under simultaneous application of both an OOP pulse and a continuous IP field. The amplitude of the OOP field pulse used for bubble expansion was kept at $H_\mathrm{z}$ = 1.9 mT, which is about 10 -- 20 \% of the depinning field \cite{HerreraDiez_APL_2015}, while its duration was varied between 1 ms and 4 ms depending on the sample. The continuous IP field $H_\mathrm{x}$ was swept between $-51.6$ mT and +51.6 mT 
\footnote{Even at the largest $H_\mathrm{x}$ applied, the velocity reached by the DW is comparable to that achieved in the presence of an $H_\mathrm{z}$ only, of magnitude up to about 30 \% of the depinning field, thus ensuring creep dynamics for the entire range of applied $H_\mathrm{x}$.}. In order to minimise OOP stray field components due to either a misalignment of the IP field or a crosstalk between IP and OOP fields, the samples were positioned carefully to ensure a symmetric growth of bubble domains for positive and negative $H_\mathrm{x}$ values. This is particularly important for magnetically soft materials like CoFeB, where $H_\mathrm{z}$ stray components in the order of 1 mT could give rise to artefacts in DW motion. 

The differential image, obtained by subtracting the expanded bubble domain image from the image of its initial state, was analysed through an image segmentation process based on the random walker algorithm \cite{Grady_2006}, which allows to precisely separate different phases in a noisy image. In our case, the two phases correspond to pixels within the area spanned by the bubble DW during its expansion or outside of it. By assigning two markers to label the `known' pixels (i.e. below and above set threshold values of grey levels), the algorithm solves a series of diffusion equations to find which of the two markers is most probable for the `unknown' pixels (whose grey levels fall between the threshold values). In this way, the initially noisy differential image was converted into a binary image from which it became possible to extract the radial profile of the initial and final state of the bubble domain. Average DW velocities could then be calculated \textit{along any desired direction} as the ratio between the DW displacement along the chosen direction and the duration time of the $H_\mathrm{z}$ field pulse.

\section{\label{sec:results1} Expansion of bubbles with Bloch and N\'eel domain walls}
In ultra-thin films with PMA, Bloch DWs are magnetostatically favoured over N\'eel ones due to the DW anisotropy field which, in the framework of the one-dimensional DW model, is written as: $H_{\mathrm{DW}} = 4 K_{\mathrm{DW}} / (\pi \mu_0 M_\mathrm{S})$, where $K_\mathrm{DW} = N_\mathrm{x} \mu_0 M_\mathrm{S}^2/2$ is the DW anisotropy energy density, $N_\mathrm{x} = \textrm{ln}(2) t / (\pi \Delta)$ is the DW demagnetising factor \cite{Tarasenko_JMMM_1998}, and $\Delta$ is the DW width. However, the interfacial DMI manifests itself as an effective in-plane field that, acting across the DW, can overcome the DW anisotropy barrier and promote a transition from Bloch to N\'eel DW texture. In the one-dimensional DW model the DMI field is expressed as \cite{Thiaville_EPL_2012}: $H_{\mathrm{DMI}} = D / \mu_0 M_\mathrm{S} \Delta$, where $D$ is the interfacial DMI constant. It follows that for $H_{\mathrm{DMI}} < H_{\mathrm{DW}}$ the DW is in a mixed Bloch-N\'eel state (purely Bloch for $D$ = 0), while a fully N\'eel DW is stabilised when $H_{\mathrm{DMI}} \geq H_{\mathrm{DW}}$.           

\begin{figure}
\includegraphics[width=1\linewidth]{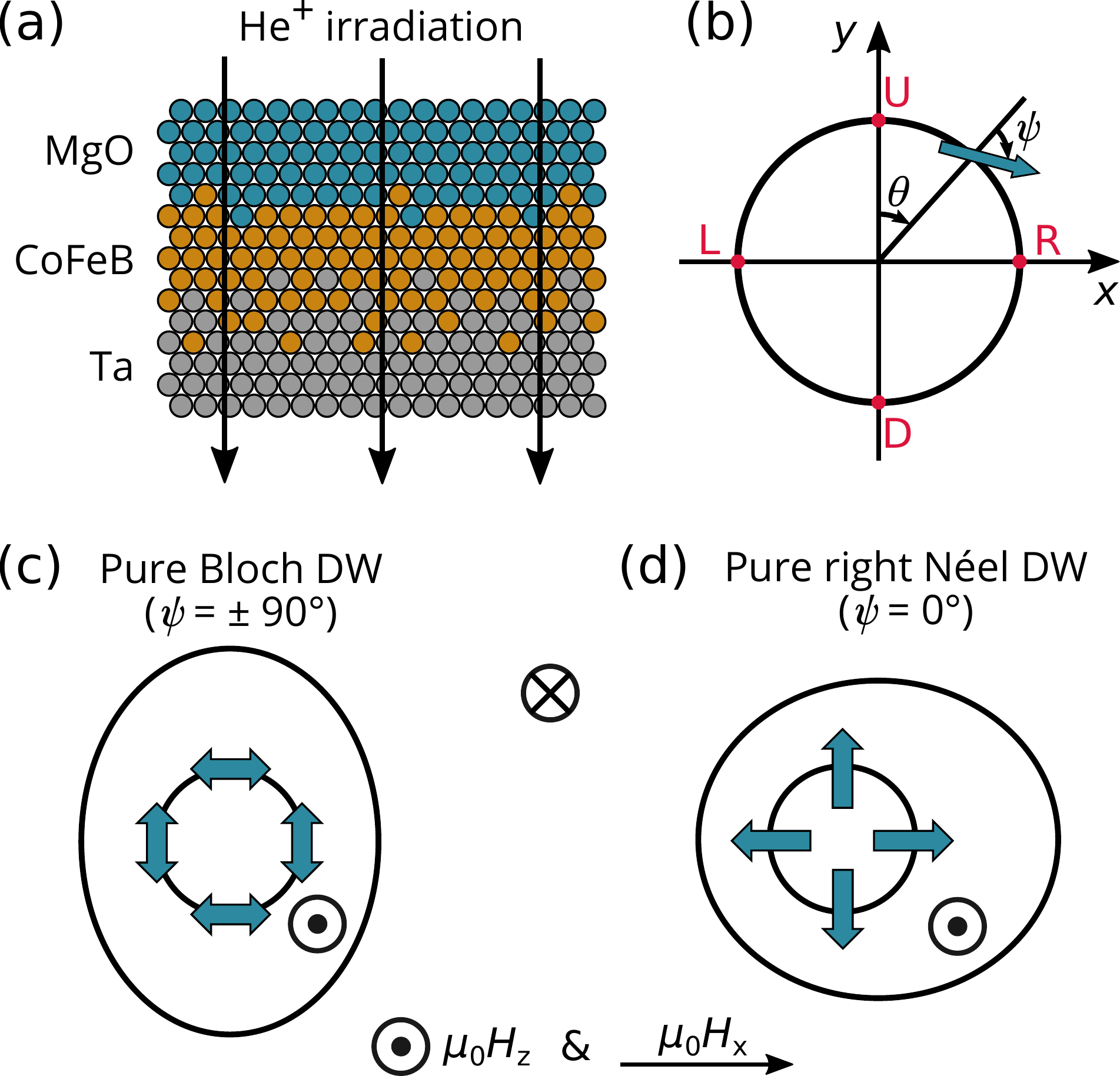}
\caption{\label{Fig1} (a) Sketch depicting the film structure and the effect of He$^+$ irradiation on the two interfaces. Irradiation-induced intermixing is more pronounced at the bottom (Ta/CoFeB) interface, while Fe depletion occurs mainly at the top (CoFeB/MgO) interface (not represented here). (b) Definition of the relevant angles and directions. $\psi$ is the magnetisation angle at the centre of the DW, measured with respect to the DW normal, while $\theta$ represents the angular position along the bubble DW. Letters ``U'', ``R'', ``D'' and ``L'' stand for up ($\theta = 0^{\circ}$), right ($\theta = 90^{\circ}$), down ($\theta = 180^{\circ}$), and left ($\theta = 270^{\circ}$) directions, respectively. Schematics of pure Bloch (c), and pure right N\'eel (d) DW structures, with corresponding bubble expansions under the simultaneous presence of IP ($H_\mathrm{x}$) and OOP ($H_\mathrm{z}$) fields. The arrows indicate the direction the magnetisation direction at the centre of the DWs.}
\end{figure}

\begin{figure*}
\includegraphics[width=0.7\linewidth]{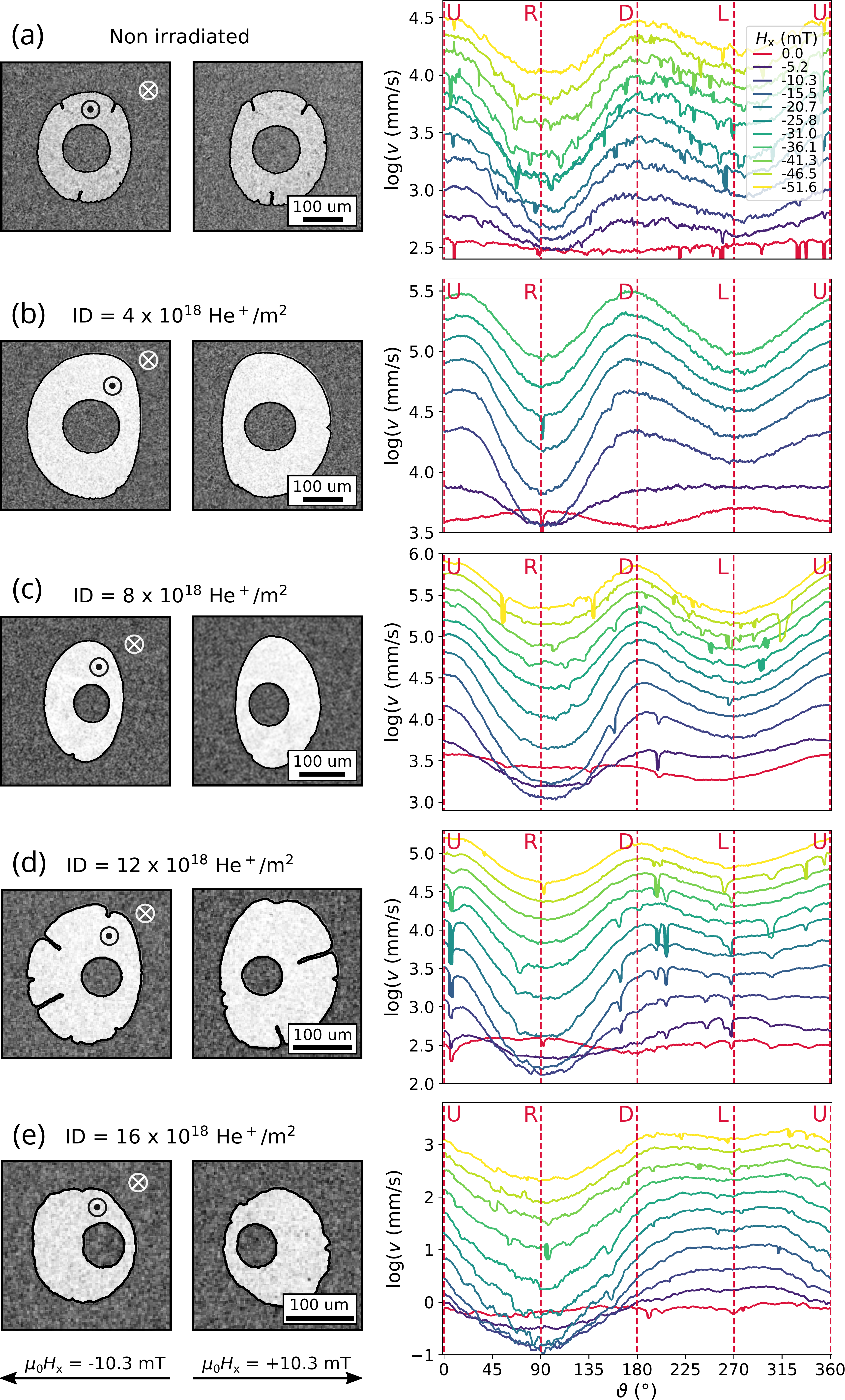}
\caption{\label{Fig2} Differential MOKE images of bubble expansion with applied $H_\mathrm{x} = \pm 10.3$ mT, and corresponding radial DW velocity profiles for increasing (negative) $H_\mathrm{x}$ values for the pristine sample (a), and the irradiated samples (b) -- (e). The applied perpendicular field is $H_\mathrm{z}$ = 1.9 mT in all cases. Letters ``U'', ``R'', ``D'' and ``L'' have been defined in Fig.~\ref{Fig1}(b). The most pronounced DW pinning points (see for instance the MOKE images in (d)) have been smoothed in the velocity profiles for better clarity.}
\end{figure*}

It has been known for almost a decade \cite{Kabanov_IEEE_2010} that magnetic bubble domains loose their initial circular shape when expanded under the simultaneous presence of IP and OOP magnetic fields. Indeed, the Zeeman energy contribution from the IP field breaks the radial symmetry of the DW energy profile, leading to DW motion at different speeds for different orientations of the internal DW magnetisation \cite{Je_PRB_2013, Hrabec_PRB_2014}. In particular, DWs move faster/slower when the magnetisation inside the DW is parallel/antiparallel to the IP field, while an intermediate velocity is found for an orthogonal alignment. This results in circular bubbles growing to acquire a wide variety of final morphologies, depending on their DW internal structure \cite{Kim_APL_2015}.

For a Bloch DW the magnetisation direction at the centre of the DW can be oriented in either of the two possible directions parallel to the DW ($\psi = \pm 90^{\circ}$), as illustrated by the double arrows in Fig.~\ref{Fig1} (c). This results in a mostly clockwise or anticlockwise arrangement of the DW magnetisation around the bubble. When both IP and OOP fields are applied, the Bloch bubble is expected to expand faster along the direction orthogonal to the IP field, while no asymmetry in the bubble shape arises along the IP field direction. Matching velocities for up and down portions of the bubble indicate identical internal DW structures, which can only arise if the clockwise/anticlockwise ordering of the DW magnetisation around the bubble is broken. This occurs when vertical Bloch lines nucleate (overcoming the local dipolar energy barrier) and give rise to head-to-head or tail-to-tail configurations within the DW \cite{Chen_NatCommun_2013, Sarma_JMMM_2018}. 

For a pure N\'eel DW the magnetisation direction inside the DW is always perpendicular to it ($\psi = 0$), with its orientation dictated by the sign of $D$ ($D > 0$ for the right-handed chirality of the N\'eel DW illustrated in Fig.~\ref{Fig1} (d)). Under application of IP and OOP fields, the N\'eel bubble expands asymmetrically along the IP field direction -- a circumstance that has been widely exploited to extract $H_{\mathrm{DMI}}$ \cite{Je_PRB_2013, Hrabec_PRB_2014} -- while the DW velocity orthogonal to the IP field is intermediate between the highest and lowest DW velocities. 

Finally, for a DW with a mixed Bloch-N\'eel character a more complex scenario is expected: in this case, the bubble should still expand asymmetrically along the IP field direction, as for a pure N\'eel bubble, but the maximum expansion should not occur in the same direction as the IP field. Despite many works reporting on the presence of mixed Bloch-N\'eel walls \cite{Hrabec_PRB_2014, Khan_APL_2016, Soucaille_PRB_2016, Wells_PRB_2017, Shepley_PRB_2018, Karnad_PRL_2018, Shahbazi_PRB_2018}, only in one case evidence has been provided of bubbles expanding with this more complex morphology \cite{Soucaille_PRB_2016}. 

\section{\label{sec:results2} Field driven DW dynamics in the creep regime}
DW motion under modest magnetic fields is explained in terms of a competition between DW elasticity and material disorder, and is phenomenologically described by the so-called creep law \cite{Lemerle_PRL_1998, Chauve_PRB_2000}:
\begin{equation}
v = v_0~\textrm{exp}{[-\zeta(\mu_0H_\mathrm{z})^{-\mu}]},
\label{eq:Creep}
\end{equation}
where $v_0$ is the characteristic speed, $\zeta$ is the scaling constant, and $\mu = 1/4$ is the creep scaling exponent. As previously explained, DW dynamics change substantially when both $H_\mathrm{z}$ and $H_\mathrm{x}$ fields are applied, but it has been proposed that the influence of $H_\mathrm{x}$ can be simply explained by changing $\zeta$ to account for the dependence of DW energy $\sigma$ on $H_\mathrm{x}$ \cite{Je_PRB_2013}:   
\begin{equation}
\zeta(H_\mathrm{x}) = \zeta_0[\sigma(H_\mathrm{x})/\sigma(0)]^{1/4},
\label{eq:Zeta}
\end{equation}
where $\zeta_0$ is another scaling constant independent of $H_\mathrm{x}$. The expression for $\sigma(H_\mathrm{x})$ along the direction of $H_\mathrm{x}$ is written differently \cite{Thiaville_EPL_2012, Je_PRB_2013}, depending on whether the DW has a mixed Bloch-N\'eel structure ($\vert$$H_\mathrm{x} + H_\mathrm{DMI}$$\vert$ $< H_\mathrm{DW}$): 
\begin{equation}
\sigma(H_\mathrm{x}) = \sigma_0 - \frac{\pi^2\Delta\mu_0^2M_\mathrm{S}^2}{8K_\mathrm{DW}}(H_\mathrm{x} + H_{\mathrm{DMI}})^2,
\label{eq:Sigma1}
\end{equation}	   
or has a purely N\'eel character ($\vert$$H_\mathrm{x} + H_\mathrm{DMI}$$\vert$ $\geq H_\mathrm{DW}$):
\begin{equation}
\sigma(H_\mathrm{x}) = \sigma_0 + 2K_\mathrm{DW}\Delta - \pi\Delta\mu_0M_\mathrm{S}|H_\mathrm{x} + H_{\mathrm{DMI}}|.
\label{eq:Sigma2}
\end{equation}
In these equations, $\sigma_0 = 4 \sqrt{AK_\mathrm{eff}}$ is the Bloch DW energy density, and $\Delta = \sqrt{A/K_\mathrm{eff}}$ is the Bloch domain wall width, with $A$ the exchange stiffness constant and $K_\mathrm{eff}$ the effective perpendicular anisotropy constant. $K_\mathrm{DW}$ is the DW anisotropy energy density that was previously defined. 

Experiments of magnetic bubble expansion in the creep regime, driven by simultaneous $H_\mathrm{z}$ and $H_\mathrm{x}$ fields, were conducted on both the pristine and the irradiated films. Fig.~\ref{Fig2} shows differential MOKE images of bubble expansion under $H_\mathrm{z} = 1.9$ mT and $H_\mathrm{x} = \pm 10.3$ mT, as well as radial DW velocity profiles for increasing (negative) IP field amplitudes up to $H_\mathrm{x} = -51.6$ mT. From the images it can be seen that the right-hand side of the bubble grows faster than the left one for positive $H_\mathrm{x}$, while the opposite is true for negative $H_\mathrm{x}$. This circumstance suggests the presence of DMI, promoting DWs with a N\'eel component of the right-hand chirality ($\uparrow \rightarrow \downarrow$ or $\downarrow \leftarrow \uparrow$) -- as it is typically found in Ta/CoFeB/MgO trilayers \cite{LoConte_PRB_2015, Khan_APL_2016, Karnad_PRL_2018, HerreraDiez_PRB_2019}. Furthermore, upon increasing ID the bubble expansion becomes progressively more asymmetric along the IP field direction, which is consistent with increasing DMI strength, as was previously reported also through Brillouin Light Scattering (BLS) measurements \cite{HerreraDiez_PRB_2019}. 
A more comprehensive analysis of the various asymmetries in the bubbles expansion is illustrated by the radial velocity curves in Fig.~\ref{Fig2}, which were extracted from the differential MOKE images for $H_\mathrm{x} < 0$ utilising an image segmentation process (see Sec.~\ref{sec:methods}). These graphs show that, while bubble domains grow rather symmetrically for $H_\mathrm{x} = 0$ mT, as expected, sizeable asymmetries start to appear when applying a modest field of $H_\mathrm{x} = -5.2$ mT. In particular, a degree of asymmetry is visible between right and left-hand sides of the bubbles, with the lowest DW velocity found around $\theta = 90^{\circ}$ for all samples, especially in the range of smaller $H_\mathrm{x}$ values. Conversely, the maximum expansion occurs predominantly along the direction perpendicular to the IP field (around $\theta = 0^{\circ}$ and $180^{\circ}$) only for samples with ID $\le 12 \times 10^{18}$ He$^+$/m$^2$ (Fig.~\ref{Fig2}(a) -- (d)), while it takes place on the left-hand side of the bubble (around $\theta = 270^{\circ}$) for the sample with highest ID (Fig.~\ref{Fig2}(e)). According to what described in section \ref{sec:results1}, these velocity behaviours indicate the presence of DWs of a mixed Bloch-N\'eel texture for films with ID $\le 12 \times 10^{18}$ He$^+$/m$^2$, whereas pure N\'eel DWs are expected for the ID = $16 \times 10^{18}$ He$^+$/m$^2$ film. In this context, it is also important to notice the extent to which the right-hand side of the bubbles is slowed down, under the application of $H_\mathrm{x}$, with respect to the expansion at $H_\mathrm{x} = 0$ mT. This information can be inferred by observing the crossing points between the velocity profile at $H_\mathrm{x} = 0$ and those at higher fields. The angular range in which the DW velocity for $H_\mathrm{x} \neq 0$ drops below that for $H_\mathrm{x} = 0$ mT becomes larger for increasing ID and progressively stronger $H_\mathrm{x}$ fields need to be applied in order to recover the velocity at $H_\mathrm{x} = 0$ mT. Since lower DW velocities correspond to higher Zeeman energies, these observations are indicative of the N\'eel component within the DW becoming larger with increasing ID \footnote{Similar conclusions can be drawn also for radial velocity curves measured under positive IP fields, which are not shown here.}. 

\begin{figure}[ht!]
\includegraphics[width=0.682\linewidth]{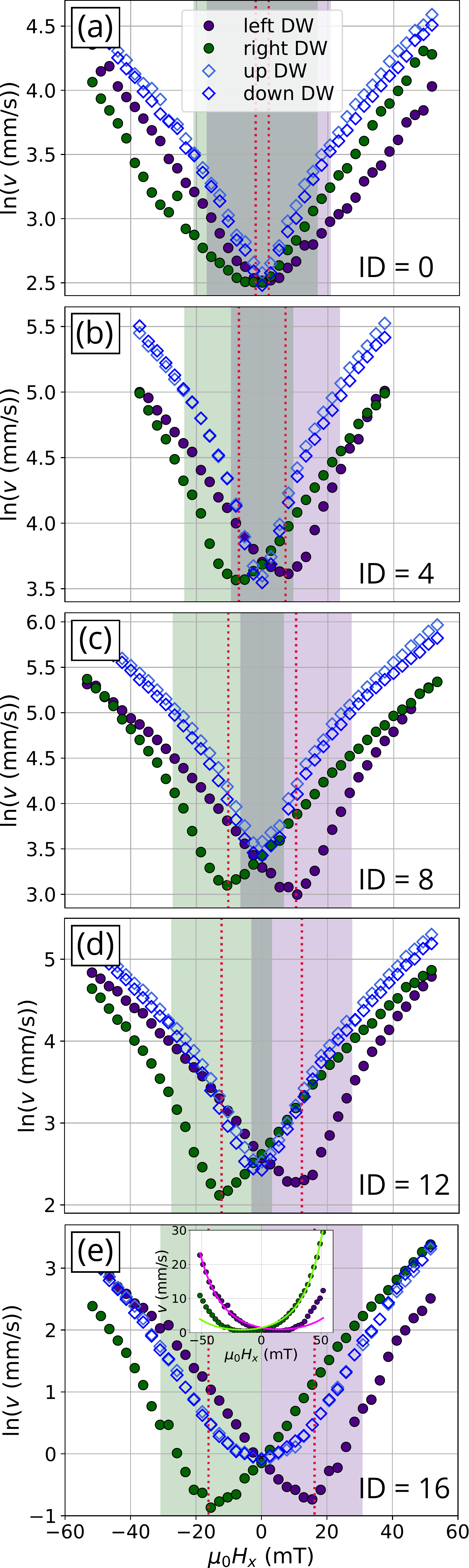}
\caption{\label{Fig3} DW velocities as a function of $H_\mathrm{x}$ for increasing ID (expressed in units of $10^{18}$ He$^+$/m$^2$). 
Green and purple areas correspond to $H_\mathrm{x}$ values for which right and left DWs, respectively, are in a mixed Bloch-N\'eel state, i.e. $\vert$$H_\mathrm{x} + H_{\mathrm{DMI}}$$\vert$ $< H_{\mathrm{DW}}$. The grey-coloured area originates from the overlapping of green and purple areas. The inset in (e) shows fitting of the left and right DW curves with Eqs.~\ref{eq:Creep} -- \ref{eq:Sigma2}.}
\end{figure}

\begin{table*}[ht]
{
\center
\renewcommand{\arraystretch}{1.5}
\begin{tabular}{c|c|c|c|c|c}
\hline 
\hline
\textbf{ID} ($\times10^{18}$ He$^+$/m$^2$) & \textit{\textbf{M$_{\mathbf{S}}$}} ($\times10^{5}$ A/m)& \textit{\textbf{K$_{\mathbf{eff}}$}} ($\times10^{5}$ J/m$^{3}$) & \textit{\textbf{A}} ($\times10^{-11}$ J/m) & $\vert$\boldmath$\mu_0$\textit{\textbf{H$_\mathbf{DMI}$}}$\vert$ (mT) & \textit{\textbf{D}} (mJ/m$^2$)\\
\hline 
\hline
0 & 8.73 & 3.45 & 2.30 & 2.0 $\pm$ 0.7 & 0.014 $\pm$ 0.005\\
\hline
4 & 7.52 & 2.66 & 1.70 & 7.1 $\pm$ 0.6 & 0.043 $\pm$ 0.004\\
\hline
8 & 8.29 & 2.77 & 2.07 & 10.3 $\pm$ 0.4 & 0.074 $\pm$ 0.003\\
\hline
12 & 7.07 & 2.28 & 1.51 & 12.2 $\pm$ 0.6 & 0.070 $\pm$ 0.003\\
\hline
16 & 6.52 & 2.08 & 1.28 & 16.2 $\pm$ 0.8 & 0.083 $\pm$ 0.04\\
\hline 
\hline
\end{tabular}
\caption{Saturation magnetisation ($M_\mathrm{S}$), effective perpendicular anisotropy constant ($K_\mathrm{eff}$), exchange stiffness constant ($A$), absolute value of the DMI field ($H_{\mathrm{DMI}}$), and DMI constant ($D$) are listed for all irradiation doses (ID). $M_\mathrm{S}$ and $K_\mathrm{eff}$ were measured by vibrating sample magnetometry (VSM) in an earlier study \cite{HerreraDiez_APL_2015}, while $A$ was estimated assuming a weak dependence on $M_\mathrm{S}^2$ (see Sec. \ref{sec:methods}), as reported in a previous work \cite{HerreraDiez_PRB_2019}. The errors in $H_{\mathrm{DMI}}$ derive from uncertainty in determining the minimum position in the DW velocity curves, and are then propagated into errors in $D$.} 
\label{Tab1}
}
\end{table*}

DW velocities along specific directions, namely $\theta = 0^{\circ}$ (up), $90^{\circ}$ (right), $180^{\circ}$ (down), and $270^{\circ}$ (left), were extracted from the complete angular profiles and plotted in Fig.~\ref{Fig3} as a function of $H_\mathrm{x}$. This more conventional way of illustrating DW velocities during bubble expansion, particularly along the direction of $H_\mathrm{x}$ (i.e. left and right DWs in the figure), is useful to extract quantitative information about the DMI strength. Indeed, according to the modified creep model \cite{Je_PRB_2013} previously outlined (see Eq.~\ref{eq:Creep} -- Eq.~\ref{eq:Sigma2}), the DW velocity curves along $H_\mathrm{x}$ should display a minimum at $H_\mathrm{x} = \pm H_{\mathrm{DMI}}$, i.e. when the DW is in the Bloch configuration. The observed shift of the velocity minimum towards higher values of $\vert$$H_\mathrm{x}$$\vert$ as a function of ID is indicative of what already mentioned: an overall increase of $D$ with increasing ID, as reported in Tab.~\ref{Tab1}. This enhancing of DMI strength through ion irradiation was previously observed and attributed to irradiation-induced intermixing at the Ta/CoFeB interface, which was revealed through x-ray reflectivity by a marked widening of the interface \cite{HerreraDiez_PRB_2019}. 
It is interesting to notice that the value of $D$ measured at the highest irradiation dose is the largest ever reported in the literature for a Ta/CoFeB/MgO trilayer, indicating the profound influence that interface disorder can have on tuning intrinsic material properties of ultra-thin films. 

The velocity curves for left and right DWs of Fig.~\ref{Fig3} could not be fitted with equations Eq.~\ref{eq:Creep} -- Eq.~\ref{eq:Sigma2} for films with ID $\le 12 \times 10^{18}$ He$^+$/m$^2$, suggesting that the dependence of $\sigma$ on $H_\mathrm{x}$ may not be the only modification to the creep law that needs to be considered to explain the influence of $H_\mathrm{x}$ on bubble expansion in these samples. Some studies have proposed that other effects may be at play, such as a domain width dependence on $H_\mathrm{x}$ \cite{Kim_APEX_2016, Sarma_JMMM_2018}, DW stiffness \cite{Pellegren_PRL_2017}, or a dependence of the DW depinning field on $H_\mathrm{x}$ \cite{Shahbazi_PRB_2019}. Interestingly, the data presented in all these works share some features with our left and right velocity curves, such as asymmetry about the minimum velocity and matching velocities in the high field region. Only for ID = 16 $\times 10^{18}$ He$^+$/m$^2$ the original modified creep model could be applied with reasonable success to fit the velocity curves, albeit not for the whole range of $H_\mathrm{x}$ values (see inset of Fig.~\ref{Fig3}(e)).
The value of $H_\mathrm{DMI}$ in this case was extracted from the fit, and was found to be in close agreement with the position of the velocity minimum. 

While the model fails to explain quantitatively some of the data in the figure, it seems to predict correctly the transition in field from mixed Bloch-N\'eel to purely N\'eel spin structure for left and right DWs. This transition is usually identified by a change of slope in the velocity curves that stems from the DW energy density $\sigma$ following Eq.~\ref{eq:Sigma1} for the mixed Bloch-N\'eel case and Eq.~\ref{eq:Sigma2} for the fully N\'eel one. The green (right DW) and purple (left DW) areas in Fig.~\ref{Fig3} correspond to field values for which $\vert$$H_\mathrm{x} + H_{\mathrm{DMI}}$$\vert$ $< H_{\mathrm{DW}}$, i.e. the mixed Bloch-N\'eel DW scenario, and a slight change of gradient in velocity is indeed evinced at the edges of these areas.   

Regarding DWs in the direction perpendicular to the IP field (up and down DWs), the velocity curves in Fig. \ref{Fig3} appear to be symmetric around $H_\mathrm{x} = 0$ and equal in magnitude, thus indicating the same expansion for up and down DWs, which is only possible with the creation of vertical Bloch lines in the case of non fully N\'eel DWs, as discussed in Sec.~\ref{sec:results1}. Furthermore, as already mentioned for Fig.~\ref{Fig2}, up and down velocities are mostly larger than left and right ones for ID $\le 12 \times 10^{18}$ He$^+$/m$^2$, while the situation reverses for ID = $16 \times 10^{18}$ He$^+$/m$^2$. This will be discussed more in the following section.

\section{\label{sec:results3} Domain wall evolution from Bloch to N\'eel structure}
Ta/CoFeB/MgO trilayers have been widely studied in the literature and, due to the usually low DMI strength, DWs are either found in a pure Bloch configuration \cite{DuttaGupta_AIPAdv_2017} or in a mixed Bloch-N\'eel one \cite{LoConte_PRB_2015, Khan_APL_2016, Karnad_PRL_2018, Ma_PRL_2018}. The results discussed so far have already given some evidence of the fact that enhancing the DMI strength of these systems through ion irradiation can promote DWs with an increasing N\'eel texture, and in particular a transition from a mostly Bloch towards a fully N\'eel spin structure. In this section further indications of DW tuning with irradiation will be provided.  

\begin{figure}[ht]
\includegraphics[width=0.8\linewidth]{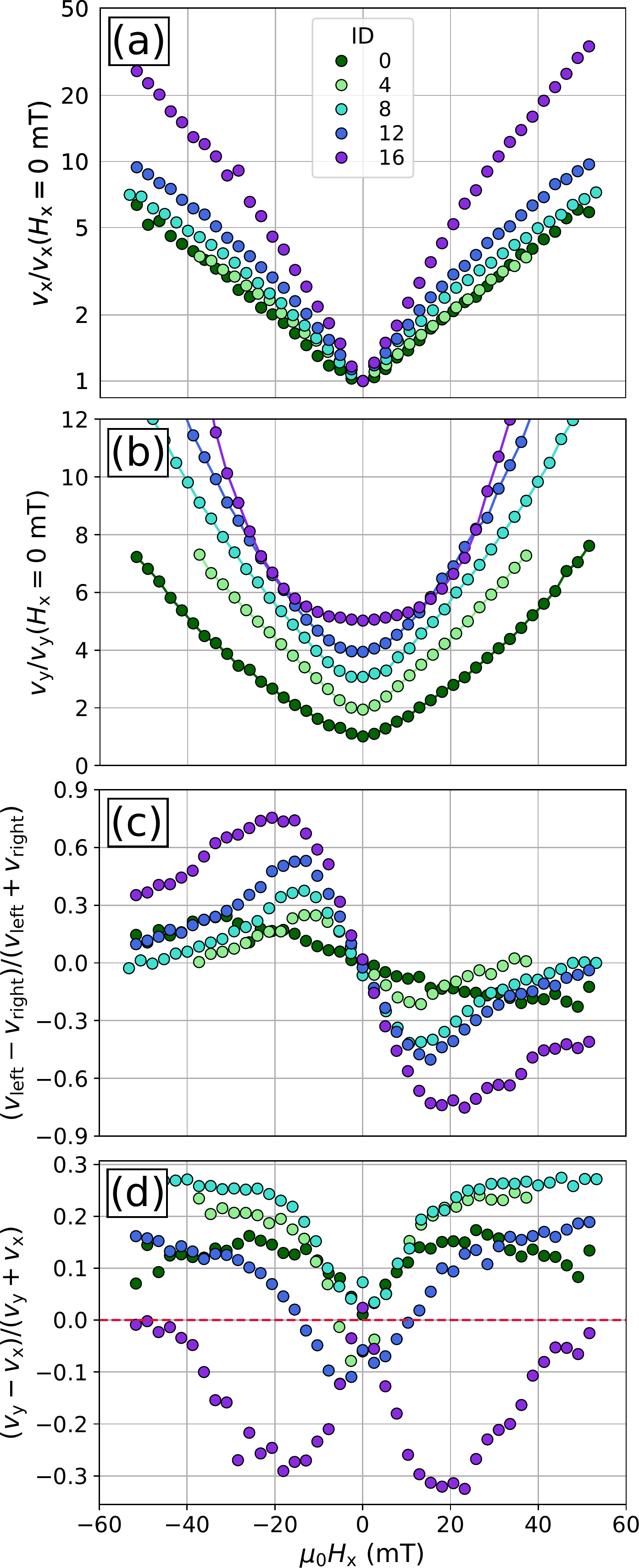}
\caption{\label{Fig4} Dependence of various quantities on the applied field $H_\mathrm{x}$ for increasing ID (expressed in units of $10^{18}$ He$^+$/m$^2$). $v_\mathrm{x} = v_\mathrm{left}$ for $H_\mathrm{x} < 0$ and $v_\mathrm{x} = v_\mathrm{right}$ for $H_\mathrm{x} > 0$, while $v_\mathrm{y}$ is the average between $v_\mathrm{up}$ and $v_\mathrm{down}$. In (b) the velocity curves for the irradiated samples have been shifted vertically for better clarity. In (d) the dashed line marks the transition between DWs with mixed Bloch-N\'eel structure and a pure N\'eel one.}
\end{figure}

Fig.~\ref{Fig4}(a) shows the ratio $v_\mathrm{x}/v_\mathrm{x}(H_\mathrm{x}=0)$ as a function of $H_\mathrm{x}$, where $v_\mathrm{x}$ coincides with the maximum velocity between left and right DWs. In other words: $v_\mathrm{x} = v_\mathrm{left}$ for $H_\mathrm{x} < 0$ and $v_\mathrm{x} = v_\mathrm{right}$ for $H_\mathrm{x} > 0$. Since DWs with internal magnetisation aligned along an applied IP field move faster than in the absence of such field, the fact that the ratio $v_\mathrm{x}/v_\mathrm{x}(H_\mathrm{x}=0)$ increases with ID is an indication of an increasing N\'eel component within the DWs. Also interesting is to analyse the same quantity along the up-down direction, namely $v_\mathrm{y}/v_\mathrm{y}(H_\mathrm{x}=0)$, which is plotted in Fig.~\ref{Fig4}(b). Here $v_\mathrm{y}$ is calculated as the average between $v_\mathrm{up}$ and $v_\mathrm{down}$. A vertical offset has been added to the velocity curves of the irradiated samples for better visualisation.	It can be seen that the curvature at the apex of these velocity curves becomes increasingly more rounded upon increasing ID -- a feature that can also be explained in terms of a Bloch-to-N\'eel transition. As a matter of fact, $H_\mathrm{x}$ favours a Bloch configuration for up and down DWs, since it is applied parallel to them. Therefore, the velocity cusp near $H_\mathrm{x} = 0$ is sharper for DWs that are already in a Bloch state or close to it (i.e. for vanishing DMI, like in the pristine sample) and becomes progressively flatter for DWs that have a stronger N\'eel component, since higher values of $H_\mathrm{x}$ are required to transition towards a Bloch DW. 

Fig.~\ref{Fig4}(c) plots the variation of bubble asymmetry along the $H_\mathrm{x}$ direction, calculated as the ratio between $v_\mathrm{left} - v_\mathrm{right}$ and $v_\mathrm{left} + v_\mathrm{right}$. Aside from the pristine sample, where the asymmetry is small, in all other cases the initial steep increase of asymmetry in modest fields is followed by a slower decline at larger fields, as was also observed in \cite{Shahbazi_PRB_2019}. 
It is interesting to notice that the amplitude of the asymmetry peak increases with ID, while its position shifts towards larger values of $\vert$$H_\mathrm{x}$$\vert$. These observations are also consistent with the scenario of a transition towards more N\'eel-like DWs with ion irradiation, since the left-right DW asymmetry is expected to be more pronounced in such case, and to reach its highest point at larger fields, which are needed to convert one of the two DWs towards a Bloch state. In this respect, the shift of the asymmetry peak towards higher fields is somewhat analogous to the displacement of the minimum velocity towards higher fields (see Fig.~\ref{Fig3}), although the two do not exactly occur at the same field.   

Finally, the behaviour of the bubble asymmetry between $x$ and $y$ directions, defined as $(v_\mathrm{y} - v_\mathrm{x})/(v_\mathrm{y} + v_\mathrm{x})$, is displayed in Fig.~\ref{Fig4}(d). There is a clear distinction in the curves of samples with ID $\leq$ 12 $\times$ 10$^{18}$ He$^+$/m$^2$, for which bubble expansion is more pronounced along the $y$ direction, and the most irradiated sample, where instead bubbles grow predominantly along $x$ for all values of applied field. As explained in section \ref{sec:results1}, this different $x$ -- $y$ asymmetry trend suggests the presence of mixed Bloch-N\'eel DWs for ID $\leq$ 12 $\times$ 10$^{18}$ He$^+$/m$^2$ and fully N\'eel DWs for the highest ID. Furthermore, for the sample with ID = 12 $\times$ 10$^{18}$ He$^+$/m$^2$ a crossover of asymmetry between $x$ and $y$ directions occurs around $\vert$$H_\mathrm{x}$$\vert$ $\sim$ 10 -- 11 mT, indicating that this sample could be the closest to the transition towards a fully N\'eel DW. These observations are consistent with the scenario of an irradiation-induced enhancement of DW N\'eel chirality. On the other hand, the fact that larger $x$ -- $y$ asymmetries are measured for ID = 4 and 8 $\times$ 10$^{18}$ He$^+$/m$^2$ rather than for the pristine sample would seemingly contradict the picture of a reduction of Bloch component upon increasing ID. However, this could be understood by taking into account the DW surface tension which plays a crucial role in the creep regime \cite{Zhang_PRA_2018}. Namely, the limited bubble growth along $x$ for the pristine sample (due to the small DMI) could constrain its growth along $y$, as the DW tension tries to maintain a circular bubble shape. Similarly, also localised pinning points (see the MOKE images in Fig. \ref{Fig2}(a)) could prevent the bubble from reaching its full expansion along $y$.  
 
\begin{figure}[ht]
\includegraphics[width=0.8\linewidth]{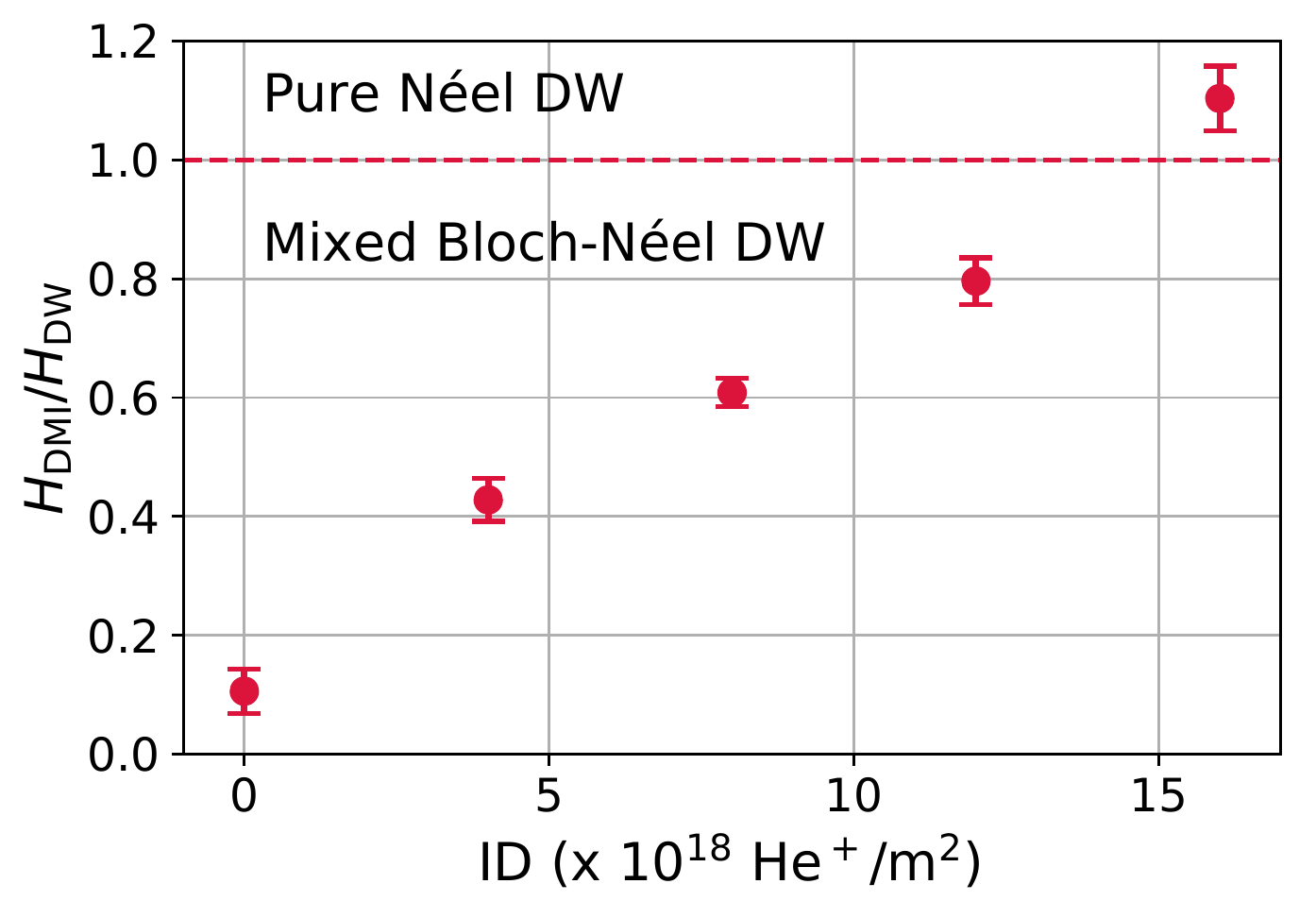}
\caption{\label{Fig5} Ratio of DMI field ($H_\mathrm{DMI}$) to DW anisotropy field ($H_\mathrm{DW}$) as a function of sample irradiation dose (ID), calculated using the parameters in Tab. \ref{Tab1}. The red dashed line marks the threshold between DWs with a mixed Bloch-N\'eel character and a fully N\'eel spin structure. The error bars derive from uncertainty in determining $H_\mathrm{DMI}$.}
\end{figure}

As was outlined in Sec.~\ref{sec:results1}, the one-dimensional DW model predicts that DWs with a purely N\'eel structure can be stabilised only if $H_\mathrm{DMI} \geq H_\mathrm{DW}$, while they retain some Bloch character if $H_\mathrm{DMI} < H_\mathrm{DW}$. In our case, the ratio $H_\mathrm{DMI}/H_\mathrm{DW}$ is found to increase approximately linearly with sample irradiation dose, as shown in Fig. \ref{Fig5}, which is in line with the qualitative trends reported so far of an enhancement of the DW N\'eel component with irradiation. Interestingly, it is only for the film with ID = 16 $\times$ 10$^{18}$ He$^+$/m$^2$ that $H_\mathrm{DMI} > H_\mathrm{DW}$, thus confirming that the different morphology of bubble expansion observed in this sample can be correlated to the presence of DWs with purely N\'eel spin structure. Since the ratio $H_\mathrm{DMI} / H_\mathrm{DW}$ is proportional to $D/M_\mathrm{S}^2$, the evolution from a Bloch to a N\'eel DW is enabled by the combined tuning of DMI strength and $M_\mathrm{S}$ through ion irradiation. 
In this respect it is important to mention that, albeit being both a direct consequence of He$^+$ irradiation, the reduction of $M_\mathrm{S}$ and the increase in DMI strength arise from distinct processes taking places at the two different interfaces. Specifically, the decrease in $M_\mathrm{S}$ is due to a depletion of Fe content at the top CoFeB/MgO interface \cite{HerreraDiez_APL_2015}, where the perpendicular magnetic anisotropy originates \cite{Ikeda_NatMater_2010}. Conversely, the increase in DMI strength is primarily caused by a widening of the bottom Ta/CoFeB interface which in turns is induced by interface intermixing \cite{HerreraDiez_PRB_2019}. This intermixing is less pronounced at the top interface, since the interfacial enthalpy promotes diffusion of Fe and Co atoms towards Ta, rather than MgO.

\section{\label{sec:conclusions} Conclusions}
To conclude, we have investigated magnetic bubble domain expansion in the creep regime under simultaneous application of IP and OOP fields for a series of Ta/CoFeB/MgO samples that were exposed to different doses of He$^+$ ion irradiation. Through the use of an image segmentation algorithm DW velocity curves were extracted for all angles along the bubble contour, which enabled us to analyse in detail the different asymmetries that arise during bubble expansion. In particular, it was shown that the overall bubble shape during growth changes upon increasing irradiation dose, with the direction of maximum expansion becoming increasingly more pronounced along the IP field direction, which can be correlated to a Bloch-to-N\'eel DW transition. Furthermore, both the DW velocity along the IP field and perpendicular to it were found to display a trend with IP field which is also consistent with the scenario of an increasing N\'eel component within the internal DW structure. These observations were correlated to two independent phenomena, both caused by He$^+$ ion irradiation: (i) atom intermixing at the Ta/CoFeB interface which results in an enhanced DMI strength, and (ii) Fe depletion at the CoFeB/MgO interface, which instead gives rise to a reduction in $M_\mathrm{S}$. Indeed, the increasing ratio $D/M_\mathrm{S}^2$ with ion irradiation is responsible for the observed tuning of DW chirality. The threshold irradiation dose above which DWs are in a purely N\'eel state was found to be between 12 $\times$ 10$^{18}$ He$^+$/m$^2$ and 16 $\times$ 10$^{18}$ He$^+$/m$^2$. Our results indicate the possibility to infer information on the DW spin texture by inspecting the overall bubble shape during evolution, and not only along the in-plane field direction. Furthermore we  show the potential of ion irradiation as a means to engineer systems with finely tuned properties, in this case enabling a transformation of the internal DW structure.

\begin{acknowledgments}
We gratefully acknowledge financial support from the EMRP Joint Research Projects 15SIB06 NanoMag and 17FUN08 TOPS, as well as from the French National Research Agency though the project ELECSPIN. We also want to acknowledge funding from the project CNRS PREMAT ``Spin-Ion''. The EMRP is jointly funded by the EMRP participating countries within EURAMET and the European Union. We are grateful to Felipe Garcia-Sanchez for useful discussions and to Luca Martino for technical support.
\end{acknowledgments}

\bibliography{biblio.bib}

\begin{thebibliography}{52}%
\makeatletter
\providecommand \@ifxundefined [1]{%
 \@ifx{#1\undefined}
}%
\providecommand \@ifnum [1]{%
 \ifnum #1\expandafter \@firstoftwo
 \else \expandafter \@secondoftwo
 \fi
}%
\providecommand \@ifx [1]{%
 \ifx #1\expandafter \@firstoftwo
 \else \expandafter \@secondoftwo
 \fi
}%
\providecommand \natexlab [1]{#1}%
\providecommand \enquote  [1]{``#1''}%
\providecommand \bibnamefont  [1]{#1}%
\providecommand \bibfnamefont [1]{#1}%
\providecommand \citenamefont [1]{#1}%
\providecommand \href@noop [0]{\@secondoftwo}%
\providecommand \href [0]{\begingroup \@sanitize@url \@href}%
\providecommand \@href[1]{\@@startlink{#1}\@@href}%
\providecommand \@@href[1]{\endgroup#1\@@endlink}%
\providecommand \@sanitize@url [0]{\catcode `\\12\catcode `\$12\catcode
  `\&12\catcode `\#12\catcode `\^12\catcode `\_12\catcode `\%12\relax}%
\providecommand \@@startlink[1]{}%
\providecommand \@@endlink[0]{}%
\providecommand \url  [0]{\begingroup\@sanitize@url \@url }%
\providecommand \@url [1]{\endgroup\@href {#1}{\urlprefix }}%
\providecommand \urlprefix  [0]{URL }%
\providecommand \Eprint [0]{\href }%
\providecommand \doibase [0]{https://doi.org/}%
\providecommand \selectlanguage [0]{\@gobble}%
\providecommand \bibinfo  [0]{\@secondoftwo}%
\providecommand \bibfield  [0]{\@secondoftwo}%
\providecommand \translation [1]{[#1]}%
\providecommand \BibitemOpen [0]{}%
\providecommand \bibitemStop [0]{}%
\providecommand \bibitemNoStop [0]{.\EOS\space}%
\providecommand \EOS [0]{\spacefactor3000\relax}%
\providecommand \BibitemShut  [1]{\csname bibitem#1\endcsname}%
\let\auto@bib@innerbib\@empty
\bibitem [{\citenamefont
  {Dzyaloshinskii}(1958)}]{Dzyaloshinsky_JPhysChemSolids_1958}%
  \BibitemOpen
  \bibfield  {author} {\bibinfo {author} {\bibfnamefont {I.}~\bibnamefont
  {Dzyaloshinskii}},\ }\href@noop {} {\bibfield  {journal} {\bibinfo  {journal}
  {J. Phys. Chem. Solids}\ }\textbf {\bibinfo {volume} {4}},\ \bibinfo {pages}
  {241} (\bibinfo {year} {1958})}\BibitemShut {NoStop}%
\bibitem [{\citenamefont {Moriya}(1960)}]{Moriya_PR_1960}%
  \BibitemOpen
  \bibfield  {author} {\bibinfo {author} {\bibfnamefont {T.}~\bibnamefont
  {Moriya}},\ }\href@noop {} {\bibfield  {journal} {\bibinfo  {journal} {Phys.
  Rev.}\ }\textbf {\bibinfo {volume} {120}},\ \bibinfo {pages} {91} (\bibinfo
  {year} {1960})}\BibitemShut {NoStop}%
\bibitem [{\citenamefont {R{\"o}{\ss}ler}\ \emph {et~al.}(2006)\citenamefont
  {R{\"o}{\ss}ler}, \citenamefont {Bogdanov},\ and\ \citenamefont
  {Pfleiderer}}]{Rossler_Nature_2006}%
  \BibitemOpen
  \bibfield  {author} {\bibinfo {author} {\bibfnamefont {U.~K.}\ \bibnamefont
  {R{\"o}{\ss}ler}}, \bibinfo {author} {\bibfnamefont {A.~N.}\ \bibnamefont
  {Bogdanov}},\ and\ \bibinfo {author} {\bibfnamefont {C.}~\bibnamefont
  {Pfleiderer}},\ }\href {https://doi.org/10.1038/nature05056} {\bibfield
  {journal} {\bibinfo  {journal} {Nature}\ }\textbf {\bibinfo {volume} {442}},\
  \bibinfo {pages} {797} (\bibinfo {year} {2006})}\BibitemShut {NoStop}%
\bibitem [{\citenamefont {M\"{u}hlbauer}\ \emph {et~al.}(2009)\citenamefont
  {M\"{u}hlbauer}, \citenamefont {Binz}, \citenamefont {Jonietz}, \citenamefont
  {Pfleiderer}, \citenamefont {Rosch}, \citenamefont {Neubauer}, \citenamefont
  {Georgii},\ and\ \citenamefont {B\"{o}ni}}]{Muhlbauer_Science_2009}%
  \BibitemOpen
  \bibfield  {author} {\bibinfo {author} {\bibfnamefont {S.}~\bibnamefont
  {M\"{u}hlbauer}}, \bibinfo {author} {\bibfnamefont {B.}~\bibnamefont {Binz}},
  \bibinfo {author} {\bibfnamefont {F.}~\bibnamefont {Jonietz}}, \bibinfo
  {author} {\bibfnamefont {C.}~\bibnamefont {Pfleiderer}}, \bibinfo {author}
  {\bibfnamefont {A.}~\bibnamefont {Rosch}}, \bibinfo {author} {\bibfnamefont
  {A.}~\bibnamefont {Neubauer}}, \bibinfo {author} {\bibfnamefont
  {R.}~\bibnamefont {Georgii}},\ and\ \bibinfo {author} {\bibfnamefont
  {P.}~\bibnamefont {B\"{o}ni}},\ }\href@noop {} {\bibfield  {journal}
  {\bibinfo  {journal} {Science}\ }\textbf {\bibinfo {volume} {323}},\ \bibinfo
  {pages} {915} (\bibinfo {year} {2009})}\BibitemShut {NoStop}%
\bibitem [{\citenamefont {Yu}\ \emph {et~al.}(2010)\citenamefont {Yu},
  \citenamefont {Onose}, \citenamefont {Kanazawa}, \citenamefont {Park},
  \citenamefont {Han}, \citenamefont {Matsui}, \citenamefont {Nagaosa},\ and\
  \citenamefont {Tokura}}]{Yu_Nature_2010}%
  \BibitemOpen
  \bibfield  {author} {\bibinfo {author} {\bibfnamefont {X.~Z.}\ \bibnamefont
  {Yu}}, \bibinfo {author} {\bibfnamefont {Y.}~\bibnamefont {Onose}}, \bibinfo
  {author} {\bibfnamefont {N.}~\bibnamefont {Kanazawa}}, \bibinfo {author}
  {\bibfnamefont {J.~H.}\ \bibnamefont {Park}}, \bibinfo {author}
  {\bibfnamefont {J.~H.}\ \bibnamefont {Han}}, \bibinfo {author} {\bibfnamefont
  {Y.}~\bibnamefont {Matsui}}, \bibinfo {author} {\bibfnamefont
  {N.}~\bibnamefont {Nagaosa}},\ and\ \bibinfo {author} {\bibfnamefont
  {Y.}~\bibnamefont {Tokura}},\ }\href {https://doi.org/10.1038/nature09124}
  {\bibfield  {journal} {\bibinfo  {journal} {Nature}\ }\textbf {\bibinfo
  {volume} {465}},\ \bibinfo {pages} {901} (\bibinfo {year}
  {2010})}\BibitemShut {NoStop}%
\bibitem [{\citenamefont {Heinze}\ \emph {et~al.}(2011)\citenamefont {Heinze},
  \citenamefont {von Bergmann}, \citenamefont {Menzel}, \citenamefont {Brede},
  \citenamefont {Kubetzka}, \citenamefont {Wiesendanger}, \citenamefont
  {Bihlmayer},\ and\ \citenamefont {Bl{\"u}gel}}]{Heinze_NatPhys_2011}%
  \BibitemOpen
  \bibfield  {author} {\bibinfo {author} {\bibfnamefont {S.}~\bibnamefont
  {Heinze}}, \bibinfo {author} {\bibfnamefont {K.}~\bibnamefont {von
  Bergmann}}, \bibinfo {author} {\bibfnamefont {M.}~\bibnamefont {Menzel}},
  \bibinfo {author} {\bibfnamefont {J.}~\bibnamefont {Brede}}, \bibinfo
  {author} {\bibfnamefont {A.}~\bibnamefont {Kubetzka}}, \bibinfo {author}
  {\bibfnamefont {R.}~\bibnamefont {Wiesendanger}}, \bibinfo {author}
  {\bibfnamefont {G.}~\bibnamefont {Bihlmayer}},\ and\ \bibinfo {author}
  {\bibfnamefont {S.}~\bibnamefont {Bl{\"u}gel}},\ }\href
  {https://doi.org/10.1038/nphys2045} {\bibfield  {journal} {\bibinfo
  {journal} {Nature Physics}\ }\textbf {\bibinfo {volume} {7}},\ \bibinfo
  {pages} {713} (\bibinfo {year} {2011})}\BibitemShut {NoStop}%
\bibitem [{\citenamefont {Emori}\ \emph {et~al.}(2013)\citenamefont {Emori},
  \citenamefont {Bauer}, \citenamefont {Ahn}, \citenamefont {Martinez},\ and\
  \citenamefont {Beach}}]{Emori_NatMater_2013}%
  \BibitemOpen
  \bibfield  {author} {\bibinfo {author} {\bibfnamefont {S.}~\bibnamefont
  {Emori}}, \bibinfo {author} {\bibfnamefont {U.}~\bibnamefont {Bauer}},
  \bibinfo {author} {\bibfnamefont {S.-M.}\ \bibnamefont {Ahn}}, \bibinfo
  {author} {\bibfnamefont {E.}~\bibnamefont {Martinez}},\ and\ \bibinfo
  {author} {\bibfnamefont {G.~S.~D.}\ \bibnamefont {Beach}},\ }\href
  {https://doi.org/10.1038/nmat3675} {\bibfield  {journal} {\bibinfo  {journal}
  {Nature Materials}\ }\textbf {\bibinfo {volume} {12}},\ \bibinfo {pages}
  {611} (\bibinfo {year} {2013})}\BibitemShut {NoStop}%
\bibitem [{\citenamefont {Ryu}\ \emph {et~al.}(2013)\citenamefont {Ryu},
  \citenamefont {Thomas}, \citenamefont {Yang},\ and\ \citenamefont
  {Parkin}}]{Ryu_NatNanotechnol_2013}%
  \BibitemOpen
  \bibfield  {author} {\bibinfo {author} {\bibfnamefont {K.-S.}\ \bibnamefont
  {Ryu}}, \bibinfo {author} {\bibfnamefont {L.}~\bibnamefont {Thomas}},
  \bibinfo {author} {\bibfnamefont {S.-H.}\ \bibnamefont {Yang}},\ and\
  \bibinfo {author} {\bibfnamefont {S.}~\bibnamefont {Parkin}},\ }\href
  {https://doi.org/10.1038/nnano.2013.102} {\bibfield  {journal} {\bibinfo
  {journal} {Nature Nanotechnology}\ }\textbf {\bibinfo {volume} {8}},\
  \bibinfo {pages} {527 EP } (\bibinfo {year} {2013})},\ \bibinfo {note}
  {article}\BibitemShut {NoStop}%
\bibitem [{\citenamefont {Moreau-Luchaire}\ \emph {et~al.}(2016)\citenamefont
  {Moreau-Luchaire}, \citenamefont {Moutafis}, \citenamefont {Reyren},
  \citenamefont {Sampaio}, \citenamefont {Vaz}, \citenamefont {Van~Horne},
  \citenamefont {Bouzehouane}, \citenamefont {Garcia}, \citenamefont
  {Deranlot}, \citenamefont {Warnicke}, \citenamefont {Wohlh{\"u}ter},
  \citenamefont {George}, \citenamefont {Weigand}, \citenamefont {Raabe},
  \citenamefont {Cros},\ and\ \citenamefont
  {Fert}}]{Moreau-Luchaire_NatNanotechnol_2016}%
  \BibitemOpen
  \bibfield  {author} {\bibinfo {author} {\bibfnamefont {C.}~\bibnamefont
  {Moreau-Luchaire}}, \bibinfo {author} {\bibfnamefont {C.}~\bibnamefont
  {Moutafis}}, \bibinfo {author} {\bibfnamefont {N.}~\bibnamefont {Reyren}},
  \bibinfo {author} {\bibfnamefont {J.}~\bibnamefont {Sampaio}}, \bibinfo
  {author} {\bibfnamefont {C.~A.~F.}\ \bibnamefont {Vaz}}, \bibinfo {author}
  {\bibfnamefont {N.}~\bibnamefont {Van~Horne}}, \bibinfo {author}
  {\bibfnamefont {K.}~\bibnamefont {Bouzehouane}}, \bibinfo {author}
  {\bibfnamefont {K.}~\bibnamefont {Garcia}}, \bibinfo {author} {\bibfnamefont
  {C.}~\bibnamefont {Deranlot}}, \bibinfo {author} {\bibfnamefont
  {P.}~\bibnamefont {Warnicke}}, \bibinfo {author} {\bibfnamefont
  {P.}~\bibnamefont {Wohlh{\"u}ter}}, \bibinfo {author} {\bibfnamefont {J.-M.}\
  \bibnamefont {George}}, \bibinfo {author} {\bibfnamefont {M.}~\bibnamefont
  {Weigand}}, \bibinfo {author} {\bibfnamefont {J.}~\bibnamefont {Raabe}},
  \bibinfo {author} {\bibfnamefont {V.}~\bibnamefont {Cros}},\ and\ \bibinfo
  {author} {\bibfnamefont {A.}~\bibnamefont {Fert}},\ }\href
  {https://doi.org/10.1038/nnano.2015.313} {\bibfield  {journal} {\bibinfo
  {journal} {Nature Nanotechnology}\ }\textbf {\bibinfo {volume} {11}},\
  \bibinfo {pages} {444} (\bibinfo {year} {2016})}\BibitemShut {NoStop}%
\bibitem [{\citenamefont {Bode}\ \emph {et~al.}(2007)\citenamefont {Bode},
  \citenamefont {Heide}, \citenamefont {von Bergmann}, \citenamefont
  {Ferriani}, \citenamefont {Heinze}, \citenamefont {Bihlmayer}, \citenamefont
  {Kubetzka}, \citenamefont {Pietzsch}, \citenamefont {Bl{\"u}gel},\ and\
  \citenamefont {Wiesendanger}}]{Bode_Nature_2007}%
  \BibitemOpen
  \bibfield  {author} {\bibinfo {author} {\bibfnamefont {M.}~\bibnamefont
  {Bode}}, \bibinfo {author} {\bibfnamefont {M.}~\bibnamefont {Heide}},
  \bibinfo {author} {\bibfnamefont {K.}~\bibnamefont {von Bergmann}}, \bibinfo
  {author} {\bibfnamefont {P.}~\bibnamefont {Ferriani}}, \bibinfo {author}
  {\bibfnamefont {S.}~\bibnamefont {Heinze}}, \bibinfo {author} {\bibfnamefont
  {G.}~\bibnamefont {Bihlmayer}}, \bibinfo {author} {\bibfnamefont
  {A.}~\bibnamefont {Kubetzka}}, \bibinfo {author} {\bibfnamefont
  {O.}~\bibnamefont {Pietzsch}}, \bibinfo {author} {\bibfnamefont
  {S.}~\bibnamefont {Bl{\"u}gel}},\ and\ \bibinfo {author} {\bibfnamefont
  {R.}~\bibnamefont {Wiesendanger}},\ }\href
  {https://doi.org/10.1038/nature05802} {\bibfield  {journal} {\bibinfo
  {journal} {Nature}\ }\textbf {\bibinfo {volume} {447}},\ \bibinfo {pages}
  {190} (\bibinfo {year} {2007})}\BibitemShut {NoStop}%
\bibitem [{\citenamefont {Thiaville}\ \emph {et~al.}(2012)\citenamefont
  {Thiaville}, \citenamefont {Rohart}, \citenamefont {Ju{\'e}}, \citenamefont
  {Cros},\ and\ \citenamefont {Fert}}]{Thiaville_EPL_2012}%
  \BibitemOpen
  \bibfield  {author} {\bibinfo {author} {\bibfnamefont {A.}~\bibnamefont
  {Thiaville}}, \bibinfo {author} {\bibfnamefont {S.}~\bibnamefont {Rohart}},
  \bibinfo {author} {\bibfnamefont {{\'E}.}~\bibnamefont {Ju{\'e}}}, \bibinfo
  {author} {\bibfnamefont {V.}~\bibnamefont {Cros}},\ and\ \bibinfo {author}
  {\bibfnamefont {A.}~\bibnamefont {Fert}},\ }\href@noop {} {\bibfield
  {journal} {\bibinfo  {journal} {Europhysics Letters}\ }\textbf {\bibinfo
  {volume} {100}},\ \bibinfo {pages} {57002} (\bibinfo {year}
  {2012})}\BibitemShut {NoStop}%
\bibitem [{\citenamefont {Yang}\ \emph {et~al.}(2015)\citenamefont {Yang},
  \citenamefont {Ryu},\ and\ \citenamefont
  {Parkin}}]{Yang_NatNanotechnol_2015}%
  \BibitemOpen
  \bibfield  {author} {\bibinfo {author} {\bibfnamefont {S.-H.}\ \bibnamefont
  {Yang}}, \bibinfo {author} {\bibfnamefont {K.-S.}\ \bibnamefont {Ryu}},\ and\
  \bibinfo {author} {\bibfnamefont {S.}~\bibnamefont {Parkin}},\ }\href
  {https://doi.org/10.1038/nnano.2014.324} {\bibfield  {journal} {\bibinfo
  {journal} {Nature Nanotechnology}\ }\textbf {\bibinfo {volume} {10}},\
  \bibinfo {pages} {221} (\bibinfo {year} {2015})}\BibitemShut {NoStop}%
\bibitem [{\citenamefont {Je}\ \emph {et~al.}(2013)\citenamefont {Je},
  \citenamefont {Kim}, \citenamefont {Yoo}, \citenamefont {Min}, \citenamefont
  {Lee},\ and\ \citenamefont {Choe}}]{Je_PRB_2013}%
  \BibitemOpen
  \bibfield  {author} {\bibinfo {author} {\bibfnamefont {S.-G.}\ \bibnamefont
  {Je}}, \bibinfo {author} {\bibfnamefont {D.-H.}\ \bibnamefont {Kim}},
  \bibinfo {author} {\bibfnamefont {S.-C.}\ \bibnamefont {Yoo}}, \bibinfo
  {author} {\bibfnamefont {B.-C.}\ \bibnamefont {Min}}, \bibinfo {author}
  {\bibfnamefont {K.-J.}\ \bibnamefont {Lee}},\ and\ \bibinfo {author}
  {\bibfnamefont {S.-B.}\ \bibnamefont {Choe}},\ }\href
  {https://doi.org/10.1103/PhysRevB.88.214401} {\bibfield  {journal} {\bibinfo
  {journal} {Phys. Rev. B}\ }\textbf {\bibinfo {volume} {88}},\ \bibinfo
  {pages} {214401} (\bibinfo {year} {2013})}\BibitemShut {NoStop}%
\bibitem [{\citenamefont {Torrejon}\ \emph {et~al.}(2014)\citenamefont
  {Torrejon}, \citenamefont {Kim}, \citenamefont {Sinha}, \citenamefont
  {Mitani}, \citenamefont {Hayashi}, \citenamefont {Yamanouchi},\ and\
  \citenamefont {Ohno}}]{Torrejon_NatCommun_2014}%
  \BibitemOpen
  \bibfield  {author} {\bibinfo {author} {\bibfnamefont {J.}~\bibnamefont
  {Torrejon}}, \bibinfo {author} {\bibfnamefont {J.}~\bibnamefont {Kim}},
  \bibinfo {author} {\bibfnamefont {J.}~\bibnamefont {Sinha}}, \bibinfo
  {author} {\bibfnamefont {S.}~\bibnamefont {Mitani}}, \bibinfo {author}
  {\bibfnamefont {M.}~\bibnamefont {Hayashi}}, \bibinfo {author} {\bibfnamefont
  {M.}~\bibnamefont {Yamanouchi}},\ and\ \bibinfo {author} {\bibfnamefont
  {H.}~\bibnamefont {Ohno}},\ }\href {https://doi.org/10.1038/ncomms5655}
  {\bibfield  {journal} {\bibinfo  {journal} {Nature Communications}\ }\textbf
  {\bibinfo {volume} {5}},\ \bibinfo {pages} {4655} (\bibinfo {year}
  {2014})}\BibitemShut {NoStop}%
\bibitem [{\citenamefont {Franken}\ \emph {et~al.}(2014)\citenamefont
  {Franken}, \citenamefont {Herps}, \citenamefont {Swagten},\ and\
  \citenamefont {Koopmans}}]{Franken_SciRep_2014}%
  \BibitemOpen
  \bibfield  {author} {\bibinfo {author} {\bibfnamefont {J.~H.}\ \bibnamefont
  {Franken}}, \bibinfo {author} {\bibfnamefont {M.}~\bibnamefont {Herps}},
  \bibinfo {author} {\bibfnamefont {H.~J.~M.}\ \bibnamefont {Swagten}},\ and\
  \bibinfo {author} {\bibfnamefont {B.}~\bibnamefont {Koopmans}},\ }\href
  {https://doi.org/10.1038/srep05248} {\bibfield  {journal} {\bibinfo
  {journal} {Scientific Reports}\ }\textbf {\bibinfo {volume} {4}},\ \bibinfo
  {pages} {5248} (\bibinfo {year} {2014})}\BibitemShut {NoStop}%
\bibitem [{\citenamefont {Benitez}\ \emph {et~al.}(2015)\citenamefont
  {Benitez}, \citenamefont {Hrabec}, \citenamefont {Mihai}, \citenamefont
  {Moore}, \citenamefont {Burnell}, \citenamefont {McGrouther}, \citenamefont
  {Marrows},\ and\ \citenamefont {McVitie}}]{Benitez_NatCommun_2015}%
  \BibitemOpen
  \bibfield  {author} {\bibinfo {author} {\bibfnamefont {M.~J.}\ \bibnamefont
  {Benitez}}, \bibinfo {author} {\bibfnamefont {A.}~\bibnamefont {Hrabec}},
  \bibinfo {author} {\bibfnamefont {A.~P.}\ \bibnamefont {Mihai}}, \bibinfo
  {author} {\bibfnamefont {T.~A.}\ \bibnamefont {Moore}}, \bibinfo {author}
  {\bibfnamefont {G.}~\bibnamefont {Burnell}}, \bibinfo {author} {\bibfnamefont
  {D.}~\bibnamefont {McGrouther}}, \bibinfo {author} {\bibfnamefont {C.~H.}\
  \bibnamefont {Marrows}},\ and\ \bibinfo {author} {\bibfnamefont
  {S.}~\bibnamefont {McVitie}},\ }\href {https://doi.org/10.1038/ncomms9957}
  {\bibfield  {journal} {\bibinfo  {journal} {Nature Communications}\ }\textbf
  {\bibinfo {volume} {6}},\ \bibinfo {pages} {8957} (\bibinfo {year} {2015})},\
  \bibinfo {note} {article}\BibitemShut {NoStop}%
\bibitem [{\citenamefont {Nembach}\ \emph {et~al.}(2015)\citenamefont
  {Nembach}, \citenamefont {Shaw}, \citenamefont {Weiler}, \citenamefont
  {Ju{\'e}},\ and\ \citenamefont {Silva}}]{Nembach_NatPhys_2015}%
  \BibitemOpen
  \bibfield  {author} {\bibinfo {author} {\bibfnamefont {H.~T.}\ \bibnamefont
  {Nembach}}, \bibinfo {author} {\bibfnamefont {J.~M.}\ \bibnamefont {Shaw}},
  \bibinfo {author} {\bibfnamefont {M.}~\bibnamefont {Weiler}}, \bibinfo
  {author} {\bibfnamefont {E.}~\bibnamefont {Ju{\'e}}},\ and\ \bibinfo {author}
  {\bibfnamefont {T.~J.}\ \bibnamefont {Silva}},\ }\href
  {https://doi.org/10.1038/nphys3418} {\bibfield  {journal} {\bibinfo
  {journal} {Nature Physics}\ }\textbf {\bibinfo {volume} {11}},\ \bibinfo
  {pages} {825} (\bibinfo {year} {2015})}\BibitemShut {NoStop}%
\bibitem [{\citenamefont {Woo}\ \emph {et~al.}(2016)\citenamefont {Woo},
  \citenamefont {Litzius}, \citenamefont {Kr{\"u}ger}, \citenamefont {Im},
  \citenamefont {Caretta}, \citenamefont {Richter}, \citenamefont {Mann},
  \citenamefont {Krone}, \citenamefont {Reeve}, \citenamefont {Weigand},
  \citenamefont {Agrawal}, \citenamefont {Lemesh}, \citenamefont {Mawass},
  \citenamefont {Fischer}, \citenamefont {Kl{\"a}ui},\ and\ \citenamefont
  {Beach}}]{Woo_NatMater_2016}%
  \BibitemOpen
  \bibfield  {author} {\bibinfo {author} {\bibfnamefont {S.}~\bibnamefont
  {Woo}}, \bibinfo {author} {\bibfnamefont {K.}~\bibnamefont {Litzius}},
  \bibinfo {author} {\bibfnamefont {B.}~\bibnamefont {Kr{\"u}ger}}, \bibinfo
  {author} {\bibfnamefont {M.-Y.}\ \bibnamefont {Im}}, \bibinfo {author}
  {\bibfnamefont {L.}~\bibnamefont {Caretta}}, \bibinfo {author} {\bibfnamefont
  {K.}~\bibnamefont {Richter}}, \bibinfo {author} {\bibfnamefont
  {M.}~\bibnamefont {Mann}}, \bibinfo {author} {\bibfnamefont {A.}~\bibnamefont
  {Krone}}, \bibinfo {author} {\bibfnamefont {R.~M.}\ \bibnamefont {Reeve}},
  \bibinfo {author} {\bibfnamefont {M.}~\bibnamefont {Weigand}}, \bibinfo
  {author} {\bibfnamefont {P.}~\bibnamefont {Agrawal}}, \bibinfo {author}
  {\bibfnamefont {I.}~\bibnamefont {Lemesh}}, \bibinfo {author} {\bibfnamefont
  {M.-A.}\ \bibnamefont {Mawass}}, \bibinfo {author} {\bibfnamefont
  {P.}~\bibnamefont {Fischer}}, \bibinfo {author} {\bibfnamefont
  {M.}~\bibnamefont {Kl{\"a}ui}},\ and\ \bibinfo {author} {\bibfnamefont
  {G.~S.~D.}\ \bibnamefont {Beach}},\ }\href {https://doi.org/10.1038/nmat4593}
  {\bibfield  {journal} {\bibinfo  {journal} {Nature Materials}\ }\textbf
  {\bibinfo {volume} {15}},\ \bibinfo {pages} {501} (\bibinfo {year}
  {2016})}\BibitemShut {NoStop}%
\bibitem [{\citenamefont {Hrabec}\ \emph {et~al.}(2014)\citenamefont {Hrabec},
  \citenamefont {Porter}, \citenamefont {Wells}, \citenamefont {Benitez},
  \citenamefont {Burnell}, \citenamefont {McVitie}, \citenamefont {McGrouther},
  \citenamefont {Moore},\ and\ \citenamefont {Marrows}}]{Hrabec_PRB_2014}%
  \BibitemOpen
  \bibfield  {author} {\bibinfo {author} {\bibfnamefont {A.}~\bibnamefont
  {Hrabec}}, \bibinfo {author} {\bibfnamefont {N.~A.}\ \bibnamefont {Porter}},
  \bibinfo {author} {\bibfnamefont {A.}~\bibnamefont {Wells}}, \bibinfo
  {author} {\bibfnamefont {M.~J.}\ \bibnamefont {Benitez}}, \bibinfo {author}
  {\bibfnamefont {G.}~\bibnamefont {Burnell}}, \bibinfo {author} {\bibfnamefont
  {S.}~\bibnamefont {McVitie}}, \bibinfo {author} {\bibfnamefont
  {D.}~\bibnamefont {McGrouther}}, \bibinfo {author} {\bibfnamefont {T.~A.}\
  \bibnamefont {Moore}},\ and\ \bibinfo {author} {\bibfnamefont {C.~H.}\
  \bibnamefont {Marrows}},\ }\href {https://doi.org/10.1103/PhysRevB.90.020402}
  {\bibfield  {journal} {\bibinfo  {journal} {Phys. Rev. B}\ }\textbf {\bibinfo
  {volume} {90}},\ \bibinfo {pages} {020402} (\bibinfo {year}
  {2014})}\BibitemShut {NoStop}%
\bibitem [{\citenamefont {Lavrijsen}\ \emph {et~al.}(2015)\citenamefont
  {Lavrijsen}, \citenamefont {Hartmann}, \citenamefont {van~den Brink},
  \citenamefont {Yin}, \citenamefont {Barcones}, \citenamefont {Duine},
  \citenamefont {Verheijen}, \citenamefont {Swagten},\ and\ \citenamefont
  {Koopmans}}]{Lavrijsen_PRB_2015}%
  \BibitemOpen
  \bibfield  {author} {\bibinfo {author} {\bibfnamefont {R.}~\bibnamefont
  {Lavrijsen}}, \bibinfo {author} {\bibfnamefont {D.~M.~F.}\ \bibnamefont
  {Hartmann}}, \bibinfo {author} {\bibfnamefont {A.}~\bibnamefont {van~den
  Brink}}, \bibinfo {author} {\bibfnamefont {Y.}~\bibnamefont {Yin}}, \bibinfo
  {author} {\bibfnamefont {B.}~\bibnamefont {Barcones}}, \bibinfo {author}
  {\bibfnamefont {R.~A.}\ \bibnamefont {Duine}}, \bibinfo {author}
  {\bibfnamefont {M.~A.}\ \bibnamefont {Verheijen}}, \bibinfo {author}
  {\bibfnamefont {H.~J.~M.}\ \bibnamefont {Swagten}},\ and\ \bibinfo {author}
  {\bibfnamefont {B.}~\bibnamefont {Koopmans}},\ }\href
  {https://doi.org/10.1103/PhysRevB.91.104414} {\bibfield  {journal} {\bibinfo
  {journal} {Phys. Rev. B}\ }\textbf {\bibinfo {volume} {91}},\ \bibinfo
  {pages} {104414} (\bibinfo {year} {2015})}\BibitemShut {NoStop}%
\bibitem [{\citenamefont {Wells}\ \emph {et~al.}(2017)\citenamefont {Wells},
  \citenamefont {Shepley}, \citenamefont {Marrows},\ and\ \citenamefont
  {Moore}}]{Wells_PRB_2017}%
  \BibitemOpen
  \bibfield  {author} {\bibinfo {author} {\bibfnamefont {A.~W.~J.}\
  \bibnamefont {Wells}}, \bibinfo {author} {\bibfnamefont {P.~M.}\ \bibnamefont
  {Shepley}}, \bibinfo {author} {\bibfnamefont {C.~H.}\ \bibnamefont
  {Marrows}},\ and\ \bibinfo {author} {\bibfnamefont {T.~A.}\ \bibnamefont
  {Moore}},\ }\href {https://doi.org/10.1103/PhysRevB.95.054428} {\bibfield
  {journal} {\bibinfo  {journal} {Phys. Rev. B}\ }\textbf {\bibinfo {volume}
  {95}},\ \bibinfo {pages} {054428} (\bibinfo {year} {2017})}\BibitemShut
  {NoStop}%
\bibitem [{\citenamefont {Chen}\ \emph {et~al.}(2013)\citenamefont {Chen},
  \citenamefont {Ma}, \citenamefont {N'Diaye}, \citenamefont {Kwon},
  \citenamefont {Won}, \citenamefont {Wu},\ and\ \citenamefont
  {Schmid}}]{Chen_NatCommun_2013}%
  \BibitemOpen
  \bibfield  {author} {\bibinfo {author} {\bibfnamefont {G.}~\bibnamefont
  {Chen}}, \bibinfo {author} {\bibfnamefont {T.}~\bibnamefont {Ma}}, \bibinfo
  {author} {\bibfnamefont {A.~T.}\ \bibnamefont {N'Diaye}}, \bibinfo {author}
  {\bibfnamefont {H.}~\bibnamefont {Kwon}}, \bibinfo {author} {\bibfnamefont
  {C.}~\bibnamefont {Won}}, \bibinfo {author} {\bibfnamefont {Y.}~\bibnamefont
  {Wu}},\ and\ \bibinfo {author} {\bibfnamefont {A.~K.}\ \bibnamefont
  {Schmid}},\ }\href {https://doi.org/10.1038/ncomms3671} {\bibfield  {journal}
  {\bibinfo  {journal} {Nature Communications}\ }\textbf {\bibinfo {volume}
  {4}},\ \bibinfo {pages} {2671} (\bibinfo {year} {2013})}\BibitemShut
  {NoStop}%
\bibitem [{\citenamefont {Chen}\ and\ \citenamefont
  {Schmid}(2015)}]{Chen_AdvMater_2015}%
  \BibitemOpen
  \bibfield  {author} {\bibinfo {author} {\bibfnamefont {G.}~\bibnamefont
  {Chen}}\ and\ \bibinfo {author} {\bibfnamefont {A.~K.}\ \bibnamefont
  {Schmid}},\ }\href@noop {} {\bibfield  {journal} {\bibinfo  {journal}
  {Advanced Materials}\ }\textbf {\bibinfo {volume} {27}},\ \bibinfo {pages}
  {5738} (\bibinfo {year} {2015})}\BibitemShut {NoStop}%
\bibitem [{\citenamefont {Cao}\ \emph {et~al.}(2018)\citenamefont {Cao},
  \citenamefont {Zhang}, \citenamefont {Koopmans}, \citenamefont {Peng},
  \citenamefont {Zhang}, \citenamefont {Wang}, \citenamefont {Yan},
  \citenamefont {Yang},\ and\ \citenamefont {Zhao}}]{Cao_Nanoscale_2018}%
  \BibitemOpen
  \bibfield  {author} {\bibinfo {author} {\bibfnamefont {A.}~\bibnamefont
  {Cao}}, \bibinfo {author} {\bibfnamefont {X.}~\bibnamefont {Zhang}}, \bibinfo
  {author} {\bibfnamefont {B.}~\bibnamefont {Koopmans}}, \bibinfo {author}
  {\bibfnamefont {S.}~\bibnamefont {Peng}}, \bibinfo {author} {\bibfnamefont
  {Y.}~\bibnamefont {Zhang}}, \bibinfo {author} {\bibfnamefont
  {Z.}~\bibnamefont {Wang}}, \bibinfo {author} {\bibfnamefont {S.}~\bibnamefont
  {Yan}}, \bibinfo {author} {\bibfnamefont {H.}~\bibnamefont {Yang}},\ and\
  \bibinfo {author} {\bibfnamefont {W.}~\bibnamefont {Zhao}},\ }\href@noop {}
  {\bibfield  {journal} {\bibinfo  {journal} {Nanoscale}\ }\textbf {\bibinfo
  {volume} {10}},\ \bibinfo {pages} {12062} (\bibinfo {year}
  {2018})}\BibitemShut {NoStop}%
\bibitem [{\citenamefont {Khan}\ \emph {et~al.}(2016)\citenamefont {Khan},
  \citenamefont {Shepley}, \citenamefont {Hrabec}, \citenamefont {Wells},
  \citenamefont {Ocker}, \citenamefont {Marrows},\ and\ \citenamefont
  {Moore}}]{Khan_APL_2016}%
  \BibitemOpen
  \bibfield  {author} {\bibinfo {author} {\bibfnamefont {R.~A.}\ \bibnamefont
  {Khan}}, \bibinfo {author} {\bibfnamefont {P.~M.}\ \bibnamefont {Shepley}},
  \bibinfo {author} {\bibfnamefont {A.}~\bibnamefont {Hrabec}}, \bibinfo
  {author} {\bibfnamefont {A.~W.~J.}\ \bibnamefont {Wells}}, \bibinfo {author}
  {\bibfnamefont {B.}~\bibnamefont {Ocker}}, \bibinfo {author} {\bibfnamefont
  {C.~H.}\ \bibnamefont {Marrows}},\ and\ \bibinfo {author} {\bibfnamefont
  {T.~A.}\ \bibnamefont {Moore}},\ }\href@noop {} {\bibfield  {journal}
  {\bibinfo  {journal} {Applied Physics Letters}\ }\textbf {\bibinfo {volume}
  {109}},\ \bibinfo {pages} {132404} (\bibinfo {year} {2016})}\BibitemShut
  {NoStop}%
\bibitem [{\citenamefont {Balk}\ \emph {et~al.}(2017)\citenamefont {Balk},
  \citenamefont {Kim}, \citenamefont {Pierce}, \citenamefont {Stiles},
  \citenamefont {Unguris},\ and\ \citenamefont {Stavis}}]{Balk_PRL_2017}%
  \BibitemOpen
  \bibfield  {author} {\bibinfo {author} {\bibfnamefont {A.~L.}\ \bibnamefont
  {Balk}}, \bibinfo {author} {\bibfnamefont {K.-W.}\ \bibnamefont {Kim}},
  \bibinfo {author} {\bibfnamefont {D.~T.}\ \bibnamefont {Pierce}}, \bibinfo
  {author} {\bibfnamefont {M.~D.}\ \bibnamefont {Stiles}}, \bibinfo {author}
  {\bibfnamefont {J.}~\bibnamefont {Unguris}},\ and\ \bibinfo {author}
  {\bibfnamefont {S.~M.}\ \bibnamefont {Stavis}},\ }\href
  {https://doi.org/10.1103/PhysRevLett.119.077205} {\bibfield  {journal}
  {\bibinfo  {journal} {Phys. Rev. Lett.}\ }\textbf {\bibinfo {volume} {119}},\
  \bibinfo {pages} {077205} (\bibinfo {year} {2017})}\BibitemShut {NoStop}%
\bibitem [{\citenamefont {Herrera~Diez}\ \emph {et~al.}(2019)\citenamefont
  {Herrera~Diez}, \citenamefont {Voto}, \citenamefont {Casiraghi},
  \citenamefont {Belmeguenai}, \citenamefont {Roussign\'e}, \citenamefont
  {Durin}, \citenamefont {Lamperti}, \citenamefont {Mantovan}, \citenamefont
  {Sluka}, \citenamefont {Jeudy}, \citenamefont {Liu}, \citenamefont
  {Stashkevich}, \citenamefont {Ch\'erif}, \citenamefont {Langer},
  \citenamefont {Ocker}, \citenamefont {Lopez-Diaz},\ and\ \citenamefont
  {Ravelosona}}]{HerreraDiez_PRB_2019}%
  \BibitemOpen
  \bibfield  {author} {\bibinfo {author} {\bibfnamefont {L.}~\bibnamefont
  {Herrera~Diez}}, \bibinfo {author} {\bibfnamefont {M.}~\bibnamefont {Voto}},
  \bibinfo {author} {\bibfnamefont {A.}~\bibnamefont {Casiraghi}}, \bibinfo
  {author} {\bibfnamefont {M.}~\bibnamefont {Belmeguenai}}, \bibinfo {author}
  {\bibfnamefont {Y.}~\bibnamefont {Roussign\'e}}, \bibinfo {author}
  {\bibfnamefont {G.}~\bibnamefont {Durin}}, \bibinfo {author} {\bibfnamefont
  {A.}~\bibnamefont {Lamperti}}, \bibinfo {author} {\bibfnamefont
  {R.}~\bibnamefont {Mantovan}}, \bibinfo {author} {\bibfnamefont
  {V.}~\bibnamefont {Sluka}}, \bibinfo {author} {\bibfnamefont
  {V.}~\bibnamefont {Jeudy}}, \bibinfo {author} {\bibfnamefont {Y.~T.}\
  \bibnamefont {Liu}}, \bibinfo {author} {\bibfnamefont {A.}~\bibnamefont
  {Stashkevich}}, \bibinfo {author} {\bibfnamefont {S.~M.}\ \bibnamefont
  {Ch\'erif}}, \bibinfo {author} {\bibfnamefont {J.}~\bibnamefont {Langer}},
  \bibinfo {author} {\bibfnamefont {B.}~\bibnamefont {Ocker}}, \bibinfo
  {author} {\bibfnamefont {L.}~\bibnamefont {Lopez-Diaz}},\ and\ \bibinfo
  {author} {\bibfnamefont {D.}~\bibnamefont {Ravelosona}},\ }\href
  {https://doi.org/10.1103/PhysRevB.99.054431} {\bibfield  {journal} {\bibinfo
  {journal} {Phys. Rev. B}\ }\textbf {\bibinfo {volume} {99}},\ \bibinfo
  {pages} {054431} (\bibinfo {year} {2019})}\BibitemShut {NoStop}%
\bibitem [{\citenamefont {Herrera~Diez}\ \emph {et~al.}(2015)\citenamefont
  {Herrera~Diez}, \citenamefont {Garc{\'i}a-S{\'a}nchez}, \citenamefont {Adam},
  \citenamefont {Devolder}, \citenamefont {Eimer}, \citenamefont {El~Hadri},
  \citenamefont {Lamperti}, \citenamefont {Mantovan}, \citenamefont {Ocker},\
  and\ \citenamefont {Ravelosona}}]{HerreraDiez_APL_2015}%
  \BibitemOpen
  \bibfield  {author} {\bibinfo {author} {\bibfnamefont {L.}~\bibnamefont
  {Herrera~Diez}}, \bibinfo {author} {\bibfnamefont {F.}~\bibnamefont
  {Garc{\'i}a-S{\'a}nchez}}, \bibinfo {author} {\bibfnamefont {J.-P.}\
  \bibnamefont {Adam}}, \bibinfo {author} {\bibfnamefont {T.}~\bibnamefont
  {Devolder}}, \bibinfo {author} {\bibfnamefont {S.}~\bibnamefont {Eimer}},
  \bibinfo {author} {\bibfnamefont {M.~S.}\ \bibnamefont {El~Hadri}}, \bibinfo
  {author} {\bibfnamefont {A.}~\bibnamefont {Lamperti}}, \bibinfo {author}
  {\bibfnamefont {R.}~\bibnamefont {Mantovan}}, \bibinfo {author}
  {\bibfnamefont {B.}~\bibnamefont {Ocker}},\ and\ \bibinfo {author}
  {\bibfnamefont {D.}~\bibnamefont {Ravelosona}},\ }\href
  {https://doi.org/10.1063/1.4927204} {\bibfield  {journal} {\bibinfo
  {journal} {Applied Physics Letters}\ }\textbf {\bibinfo {volume} {107}},\
  \bibinfo {pages} {032401} (\bibinfo {year} {2015})}\BibitemShut {NoStop}%
\bibitem [{\citenamefont {Yamanouchi}\ \emph {et~al.}(2011)\citenamefont
  {Yamanouchi}, \citenamefont {Jander}, \citenamefont {Dhagat}, \citenamefont
  {Ikeda}, \citenamefont {Matsukura},\ and\ \citenamefont
  {Ohno}}]{Yamanouchi_IEEEMagnLett_2011}%
  \BibitemOpen
  \bibfield  {author} {\bibinfo {author} {\bibfnamefont {M.}~\bibnamefont
  {Yamanouchi}}, \bibinfo {author} {\bibfnamefont {A.}~\bibnamefont {Jander}},
  \bibinfo {author} {\bibfnamefont {P.}~\bibnamefont {Dhagat}}, \bibinfo
  {author} {\bibfnamefont {S.}~\bibnamefont {Ikeda}}, \bibinfo {author}
  {\bibfnamefont {F.}~\bibnamefont {Matsukura}},\ and\ \bibinfo {author}
  {\bibfnamefont {H.}~\bibnamefont {Ohno}},\ }\href
  {https://doi.org/10.1109/LMAG.2011.2159484} {\bibfield  {journal} {\bibinfo
  {journal} {IEEE Magnetics Letters}\ }\textbf {\bibinfo {volume} {2}},\
  \bibinfo {pages} {3000304} (\bibinfo {year} {2011})}\BibitemShut {NoStop}%
\bibitem [{\citenamefont {Asti}\ \emph {et~al.}(2007)\citenamefont {Asti},
  \citenamefont {Ghidini}, \citenamefont {Mulazzi}, \citenamefont {Pellicelli},
  \citenamefont {Solzi}, \citenamefont {Chesnel},\ and\ \citenamefont
  {Marty}}]{Asti_PRB_2007}%
  \BibitemOpen
  \bibfield  {author} {\bibinfo {author} {\bibfnamefont {G.}~\bibnamefont
  {Asti}}, \bibinfo {author} {\bibfnamefont {M.}~\bibnamefont {Ghidini}},
  \bibinfo {author} {\bibfnamefont {M.}~\bibnamefont {Mulazzi}}, \bibinfo
  {author} {\bibfnamefont {R.}~\bibnamefont {Pellicelli}}, \bibinfo {author}
  {\bibfnamefont {M.}~\bibnamefont {Solzi}}, \bibinfo {author} {\bibfnamefont
  {K.}~\bibnamefont {Chesnel}},\ and\ \bibinfo {author} {\bibfnamefont
  {A.}~\bibnamefont {Marty}},\ }\href
  {https://doi.org/10.1103/PhysRevB.76.094414} {\bibfield  {journal} {\bibinfo
  {journal} {Phys. Rev. B}\ }\textbf {\bibinfo {volume} {76}},\ \bibinfo
  {pages} {094414} (\bibinfo {year} {2007})}\BibitemShut {NoStop}%
\bibitem [{\citenamefont {Hubert}\ and\ \citenamefont
  {Sch\"{a}fer}(1998)}]{Hubert}%
  \BibitemOpen
  \bibfield  {author} {\bibinfo {author} {\bibfnamefont {A.}~\bibnamefont
  {Hubert}}\ and\ \bibinfo {author} {\bibfnamefont {R.}~\bibnamefont
  {Sch\"{a}fer}},\ }\bibinfo {title} {Magnetic domains}\ (\bibinfo  {publisher}
  {Springer-Verlag},\ \bibinfo {year} {1998})\ p.\ \bibinfo {pages}
  {390}\BibitemShut {NoStop}%
\bibitem [{Note1()}]{Note1}%
  \BibitemOpen
  \bibinfo {note} {Even at the largest $H_\protect \mathrm {x}$ applied, the
  velocity reached by the DW is comparable to that achieved in the presence of
  an $H_\protect \mathrm {z}$ only, of magnitude up to about 30 \% of the
  depinning field, thus ensuring creep dynamics for the entire range of applied
  $H_\protect \mathrm {x}$.}\BibitemShut {Stop}%
\bibitem [{\citenamefont {Grady}(2006)}]{Grady_2006}%
  \BibitemOpen
  \bibfield  {author} {\bibinfo {author} {\bibfnamefont {L.}~\bibnamefont
  {Grady}},\ }\href@noop {} {\bibfield  {journal} {\bibinfo  {journal} {IEEE
  Transactions on Pattern Analysis and Machine Intelligence}\ }\textbf
  {\bibinfo {volume} {28}},\ \bibinfo {pages} {1768} (\bibinfo {year}
  {2006})}\BibitemShut {NoStop}%
\bibitem [{\citenamefont {Tarasenko}\ \emph {et~al.}(1998)\citenamefont
  {Tarasenko}, \citenamefont {Stankiewicz}, \citenamefont {Tarasenko},\ and\
  \citenamefont {Ferr{\'e}}}]{Tarasenko_JMMM_1998}%
  \BibitemOpen
  \bibfield  {author} {\bibinfo {author} {\bibfnamefont {S.~V.}\ \bibnamefont
  {Tarasenko}}, \bibinfo {author} {\bibfnamefont {A.}~\bibnamefont
  {Stankiewicz}}, \bibinfo {author} {\bibfnamefont {V.~V.}\ \bibnamefont
  {Tarasenko}},\ and\ \bibinfo {author} {\bibfnamefont {J.}~\bibnamefont
  {Ferr{\'e}}},\ }\href
  {http://www.sciencedirect.com/science/article/pii/S0304885398002303}
  {\bibfield  {journal} {\bibinfo  {journal} {Journal of Magnetism and Magnetic
  Materials}\ }\textbf {\bibinfo {volume} {189}},\ \bibinfo {pages} {19}
  (\bibinfo {year} {1998})}\BibitemShut {NoStop}%
\bibitem [{\citenamefont {Kabanov}\ \emph {et~al.}(2010)\citenamefont
  {Kabanov}, \citenamefont {Iunin}, \citenamefont {Nikitenko}, \citenamefont
  {Shapiro}, \citenamefont {Shull}, \citenamefont {Zhu},\ and\ \citenamefont
  {Chien}}]{Kabanov_IEEE_2010}%
  \BibitemOpen
  \bibfield  {author} {\bibinfo {author} {\bibfnamefont {Y.~P.}\ \bibnamefont
  {Kabanov}}, \bibinfo {author} {\bibfnamefont {Y.~L.}\ \bibnamefont {Iunin}},
  \bibinfo {author} {\bibfnamefont {V.~I.}\ \bibnamefont {Nikitenko}}, \bibinfo
  {author} {\bibfnamefont {A.~J.}\ \bibnamefont {Shapiro}}, \bibinfo {author}
  {\bibfnamefont {R.~D.}\ \bibnamefont {Shull}}, \bibinfo {author}
  {\bibfnamefont {L.~Y.}\ \bibnamefont {Zhu}},\ and\ \bibinfo {author}
  {\bibfnamefont {C.~L.}\ \bibnamefont {Chien}},\ }\href
  {https://doi.org/10.1109/TMAG.2010.2045740} {\bibfield  {journal} {\bibinfo
  {journal} {IEEE Transactions on Magnetics}\ }\textbf {\bibinfo {volume}
  {46}},\ \bibinfo {pages} {2220} (\bibinfo {year} {2010})}\BibitemShut
  {NoStop}%
\bibitem [{\citenamefont {Kim}\ \emph {et~al.}(2015)\citenamefont {Kim},
  \citenamefont {Kim}, \citenamefont {Moon},\ and\ \citenamefont
  {Choe}}]{Kim_APL_2015}%
  \BibitemOpen
  \bibfield  {author} {\bibinfo {author} {\bibfnamefont {D.-Y.}\ \bibnamefont
  {Kim}}, \bibinfo {author} {\bibfnamefont {D.-H.}\ \bibnamefont {Kim}},
  \bibinfo {author} {\bibfnamefont {J.}~\bibnamefont {Moon}},\ and\ \bibinfo
  {author} {\bibfnamefont {S.-B.}\ \bibnamefont {Choe}},\ }\href
  {https://doi.org/10.1063/1.4922943} {\bibfield  {journal} {\bibinfo
  {journal} {Applied Physics Letters}\ }\textbf {\bibinfo {volume} {106}},\
  \bibinfo {pages} {262403} (\bibinfo {year} {2015})}\BibitemShut {NoStop}%
\bibitem [{\citenamefont {Sarma}\ \emph {et~al.}(2018)\citenamefont {Sarma},
  \citenamefont {Garcia-Sanchez}, \citenamefont {Nasseri}, \citenamefont
  {Casiraghi},\ and\ \citenamefont {Durin}}]{Sarma_JMMM_2018}%
  \BibitemOpen
  \bibfield  {author} {\bibinfo {author} {\bibfnamefont {B.}~\bibnamefont
  {Sarma}}, \bibinfo {author} {\bibfnamefont {F.}~\bibnamefont
  {Garcia-Sanchez}}, \bibinfo {author} {\bibfnamefont {S.~A.}\ \bibnamefont
  {Nasseri}}, \bibinfo {author} {\bibfnamefont {A.}~\bibnamefont {Casiraghi}},\
  and\ \bibinfo {author} {\bibfnamefont {G.}~\bibnamefont {Durin}},\ }\href
  {http://www.sciencedirect.com/science/article/pii/S0304885317333589}
  {\bibfield  {journal} {\bibinfo  {journal} {Journal of Magnetism and Magnetic
  Materials}\ }\textbf {\bibinfo {volume} {456}},\ \bibinfo {pages} {433}
  (\bibinfo {year} {2018})}\BibitemShut {NoStop}%
\bibitem [{\citenamefont {Soucaille}\ \emph {et~al.}(2016)\citenamefont
  {Soucaille}, \citenamefont {Belmeguenai}, \citenamefont {Torrejon},
  \citenamefont {Kim}, \citenamefont {Devolder}, \citenamefont {Roussign\'e},
  \citenamefont {Ch\'erif}, \citenamefont {Stashkevich}, \citenamefont
  {Hayashi},\ and\ \citenamefont {Adam}}]{Soucaille_PRB_2016}%
  \BibitemOpen
  \bibfield  {author} {\bibinfo {author} {\bibfnamefont {R.}~\bibnamefont
  {Soucaille}}, \bibinfo {author} {\bibfnamefont {M.}~\bibnamefont
  {Belmeguenai}}, \bibinfo {author} {\bibfnamefont {J.}~\bibnamefont
  {Torrejon}}, \bibinfo {author} {\bibfnamefont {J.-V.}\ \bibnamefont {Kim}},
  \bibinfo {author} {\bibfnamefont {T.}~\bibnamefont {Devolder}}, \bibinfo
  {author} {\bibfnamefont {Y.}~\bibnamefont {Roussign\'e}}, \bibinfo {author}
  {\bibfnamefont {S.-M.}\ \bibnamefont {Ch\'erif}}, \bibinfo {author}
  {\bibfnamefont {A.~A.}\ \bibnamefont {Stashkevich}}, \bibinfo {author}
  {\bibfnamefont {M.}~\bibnamefont {Hayashi}},\ and\ \bibinfo {author}
  {\bibfnamefont {J.-P.}\ \bibnamefont {Adam}},\ }\href@noop {} {\bibfield
  {journal} {\bibinfo  {journal} {Phys. Rev. B}\ }\textbf {\bibinfo {volume}
  {94}},\ \bibinfo {pages} {104431} (\bibinfo {year} {2016})}\BibitemShut
  {NoStop}%
\bibitem [{\citenamefont {Shepley}\ \emph {et~al.}(2018)\citenamefont
  {Shepley}, \citenamefont {Tunnicliffe}, \citenamefont {Shahbazi},
  \citenamefont {Burnell},\ and\ \citenamefont {Moore}}]{Shepley_PRB_2018}%
  \BibitemOpen
  \bibfield  {author} {\bibinfo {author} {\bibfnamefont {P.~M.}\ \bibnamefont
  {Shepley}}, \bibinfo {author} {\bibfnamefont {H.}~\bibnamefont
  {Tunnicliffe}}, \bibinfo {author} {\bibfnamefont {K.}~\bibnamefont
  {Shahbazi}}, \bibinfo {author} {\bibfnamefont {G.}~\bibnamefont {Burnell}},\
  and\ \bibinfo {author} {\bibfnamefont {T.~A.}\ \bibnamefont {Moore}},\ }\href
  {https://doi.org/10.1103/PhysRevB.97.134417} {\bibfield  {journal} {\bibinfo
  {journal} {Phys. Rev. B}\ }\textbf {\bibinfo {volume} {97}},\ \bibinfo
  {pages} {134417} (\bibinfo {year} {2018})}\BibitemShut {NoStop}%
\bibitem [{\citenamefont {Karnad}\ \emph {et~al.}(2018)\citenamefont {Karnad},
  \citenamefont {Freimuth}, \citenamefont {Martinez}, \citenamefont {Lo~Conte},
  \citenamefont {Gubbiotti}, \citenamefont {Schulz}, \citenamefont {Senz},
  \citenamefont {Ocker}, \citenamefont {Mokrousov},\ and\ \citenamefont
  {Kl\"aui}}]{Karnad_PRL_2018}%
  \BibitemOpen
  \bibfield  {author} {\bibinfo {author} {\bibfnamefont {G.~V.}\ \bibnamefont
  {Karnad}}, \bibinfo {author} {\bibfnamefont {F.}~\bibnamefont {Freimuth}},
  \bibinfo {author} {\bibfnamefont {E.}~\bibnamefont {Martinez}}, \bibinfo
  {author} {\bibfnamefont {R.}~\bibnamefont {Lo~Conte}}, \bibinfo {author}
  {\bibfnamefont {G.}~\bibnamefont {Gubbiotti}}, \bibinfo {author}
  {\bibfnamefont {T.}~\bibnamefont {Schulz}}, \bibinfo {author} {\bibfnamefont
  {S.}~\bibnamefont {Senz}}, \bibinfo {author} {\bibfnamefont {B.}~\bibnamefont
  {Ocker}}, \bibinfo {author} {\bibfnamefont {Y.}~\bibnamefont {Mokrousov}},\
  and\ \bibinfo {author} {\bibfnamefont {M.}~\bibnamefont {Kl\"aui}},\ }\href
  {https://doi.org/10.1103/PhysRevLett.121.147203} {\bibfield  {journal}
  {\bibinfo  {journal} {Phys. Rev. Lett.}\ }\textbf {\bibinfo {volume} {121}},\
  \bibinfo {pages} {147203} (\bibinfo {year} {2018})}\BibitemShut {NoStop}%
\bibitem [{\citenamefont {Shahbazi}\ \emph {et~al.}(2018)\citenamefont
  {Shahbazi}, \citenamefont {Hrabec}, \citenamefont {Moretti}, \citenamefont
  {Ward}, \citenamefont {Moore}, \citenamefont {Jeudy}, \citenamefont
  {Martinez},\ and\ \citenamefont {Marrows}}]{Shahbazi_PRB_2018}%
  \BibitemOpen
  \bibfield  {author} {\bibinfo {author} {\bibfnamefont {K.}~\bibnamefont
  {Shahbazi}}, \bibinfo {author} {\bibfnamefont {A.~c.~v.}\ \bibnamefont
  {Hrabec}}, \bibinfo {author} {\bibfnamefont {S.}~\bibnamefont {Moretti}},
  \bibinfo {author} {\bibfnamefont {M.~B.}\ \bibnamefont {Ward}}, \bibinfo
  {author} {\bibfnamefont {T.~A.}\ \bibnamefont {Moore}}, \bibinfo {author}
  {\bibfnamefont {V.}~\bibnamefont {Jeudy}}, \bibinfo {author} {\bibfnamefont
  {E.}~\bibnamefont {Martinez}},\ and\ \bibinfo {author} {\bibfnamefont
  {C.~H.}\ \bibnamefont {Marrows}},\ }\href
  {https://doi.org/10.1103/PhysRevB.98.214413} {\bibfield  {journal} {\bibinfo
  {journal} {Phys. Rev. B}\ }\textbf {\bibinfo {volume} {98}},\ \bibinfo
  {pages} {214413} (\bibinfo {year} {2018})}\BibitemShut {NoStop}%
\bibitem [{\citenamefont {Lemerle}\ \emph {et~al.}(1998)\citenamefont
  {Lemerle}, \citenamefont {Ferr\'e}, \citenamefont {Chappert}, \citenamefont
  {Mathet}, \citenamefont {Giamarchi},\ and\ \citenamefont
  {Le~Doussal}}]{Lemerle_PRL_1998}%
  \BibitemOpen
  \bibfield  {author} {\bibinfo {author} {\bibfnamefont {S.}~\bibnamefont
  {Lemerle}}, \bibinfo {author} {\bibfnamefont {J.}~\bibnamefont {Ferr\'e}},
  \bibinfo {author} {\bibfnamefont {C.}~\bibnamefont {Chappert}}, \bibinfo
  {author} {\bibfnamefont {V.}~\bibnamefont {Mathet}}, \bibinfo {author}
  {\bibfnamefont {T.}~\bibnamefont {Giamarchi}},\ and\ \bibinfo {author}
  {\bibfnamefont {P.}~\bibnamefont {Le~Doussal}},\ }\href
  {https://doi.org/10.1103/PhysRevLett.80.849} {\bibfield  {journal} {\bibinfo
  {journal} {Phys. Rev. Lett.}\ }\textbf {\bibinfo {volume} {80}},\ \bibinfo
  {pages} {849} (\bibinfo {year} {1998})}\BibitemShut {NoStop}%
\bibitem [{\citenamefont {Chauve}\ \emph {et~al.}(2000)\citenamefont {Chauve},
  \citenamefont {Giamarchi},\ and\ \citenamefont
  {Le~Doussal}}]{Chauve_PRB_2000}%
  \BibitemOpen
  \bibfield  {author} {\bibinfo {author} {\bibfnamefont {P.}~\bibnamefont
  {Chauve}}, \bibinfo {author} {\bibfnamefont {T.}~\bibnamefont {Giamarchi}},\
  and\ \bibinfo {author} {\bibfnamefont {P.}~\bibnamefont {Le~Doussal}},\
  }\href {https://doi.org/10.1103/PhysRevB.62.6241} {\bibfield  {journal}
  {\bibinfo  {journal} {Phys. Rev. B}\ }\textbf {\bibinfo {volume} {62}},\
  \bibinfo {pages} {6241} (\bibinfo {year} {2000})}\BibitemShut {NoStop}%
\bibitem [{\citenamefont {Lo~Conte}\ \emph {et~al.}(2015)\citenamefont
  {Lo~Conte}, \citenamefont {Martinez}, \citenamefont {Hrabec}, \citenamefont
  {Lamperti}, \citenamefont {Schulz}, \citenamefont {Nasi}, \citenamefont
  {Lazzarini}, \citenamefont {Mantovan}, \citenamefont {Maccherozzi},
  \citenamefont {Dhesi}, \citenamefont {Ocker}, \citenamefont {Marrows},
  \citenamefont {Moore},\ and\ \citenamefont {Kl\"aui}}]{LoConte_PRB_2015}%
  \BibitemOpen
  \bibfield  {author} {\bibinfo {author} {\bibfnamefont {R.}~\bibnamefont
  {Lo~Conte}}, \bibinfo {author} {\bibfnamefont {E.}~\bibnamefont {Martinez}},
  \bibinfo {author} {\bibfnamefont {A.}~\bibnamefont {Hrabec}}, \bibinfo
  {author} {\bibfnamefont {A.}~\bibnamefont {Lamperti}}, \bibinfo {author}
  {\bibfnamefont {T.}~\bibnamefont {Schulz}}, \bibinfo {author} {\bibfnamefont
  {L.}~\bibnamefont {Nasi}}, \bibinfo {author} {\bibfnamefont {L.}~\bibnamefont
  {Lazzarini}}, \bibinfo {author} {\bibfnamefont {R.}~\bibnamefont {Mantovan}},
  \bibinfo {author} {\bibfnamefont {F.}~\bibnamefont {Maccherozzi}}, \bibinfo
  {author} {\bibfnamefont {S.~S.}\ \bibnamefont {Dhesi}}, \bibinfo {author}
  {\bibfnamefont {B.}~\bibnamefont {Ocker}}, \bibinfo {author} {\bibfnamefont
  {C.~H.}\ \bibnamefont {Marrows}}, \bibinfo {author} {\bibfnamefont {T.~A.}\
  \bibnamefont {Moore}},\ and\ \bibinfo {author} {\bibfnamefont
  {M.}~\bibnamefont {Kl\"aui}},\ }\href@noop {} {\bibfield  {journal} {\bibinfo
   {journal} {Phys. Rev. B}\ }\textbf {\bibinfo {volume} {91}},\ \bibinfo
  {pages} {014433} (\bibinfo {year} {2015})}\BibitemShut {NoStop}%
\bibitem [{Note2()}]{Note2}%
  \BibitemOpen
  \bibinfo {note} {Similar conclusions can be drawn also for radial velocity
  curves measured under positive IP fields, which are not shown
  here.}\BibitemShut {Stop}%
\bibitem [{\citenamefont {Kim}\ \emph {et~al.}(2016)\citenamefont {Kim},
  \citenamefont {Kim},\ and\ \citenamefont {Choe}}]{Kim_APEX_2016}%
  \BibitemOpen
  \bibfield  {author} {\bibinfo {author} {\bibfnamefont {D.-Y.}\ \bibnamefont
  {Kim}}, \bibinfo {author} {\bibfnamefont {D.-H.}\ \bibnamefont {Kim}},\ and\
  \bibinfo {author} {\bibfnamefont {S.-B.}\ \bibnamefont {Choe}},\ }\href
  {https://doi.org/10.7567/apex.9.053001} {\ \textbf {\bibinfo {volume} {9}},\
  \bibinfo {pages} {053001} (\bibinfo {year} {2016})}\BibitemShut {NoStop}%
\bibitem [{\citenamefont {Pellegren}\ \emph {et~al.}(2017)\citenamefont
  {Pellegren}, \citenamefont {Lau},\ and\ \citenamefont
  {Sokalski}}]{Pellegren_PRL_2017}%
  \BibitemOpen
  \bibfield  {author} {\bibinfo {author} {\bibfnamefont {J.~P.}\ \bibnamefont
  {Pellegren}}, \bibinfo {author} {\bibfnamefont {D.}~\bibnamefont {Lau}},\
  and\ \bibinfo {author} {\bibfnamefont {V.}~\bibnamefont {Sokalski}},\ }\href
  {https://doi.org/10.1103/PhysRevLett.119.027203} {\bibfield  {journal}
  {\bibinfo  {journal} {Phys. Rev. Lett.}\ }\textbf {\bibinfo {volume} {119}},\
  \bibinfo {pages} {027203} (\bibinfo {year} {2017})}\BibitemShut {NoStop}%
\bibitem [{\citenamefont {Shahbazi}\ \emph {et~al.}(2019)\citenamefont
  {Shahbazi}, \citenamefont {Kim}, \citenamefont {Nembach}, \citenamefont
  {Shaw}, \citenamefont {Bischof}, \citenamefont {Rossell}, \citenamefont
  {Jeudy}, \citenamefont {Moore},\ and\ \citenamefont
  {Marrows}}]{Shahbazi_PRB_2019}%
  \BibitemOpen
  \bibfield  {author} {\bibinfo {author} {\bibfnamefont {K.}~\bibnamefont
  {Shahbazi}}, \bibinfo {author} {\bibfnamefont {J.-V.}\ \bibnamefont {Kim}},
  \bibinfo {author} {\bibfnamefont {H.~T.}\ \bibnamefont {Nembach}}, \bibinfo
  {author} {\bibfnamefont {J.~M.}\ \bibnamefont {Shaw}}, \bibinfo {author}
  {\bibfnamefont {A.}~\bibnamefont {Bischof}}, \bibinfo {author} {\bibfnamefont
  {M.~D.}\ \bibnamefont {Rossell}}, \bibinfo {author} {\bibfnamefont
  {V.}~\bibnamefont {Jeudy}}, \bibinfo {author} {\bibfnamefont {T.~A.}\
  \bibnamefont {Moore}},\ and\ \bibinfo {author} {\bibfnamefont {C.~H.}\
  \bibnamefont {Marrows}},\ }\href {https://doi.org/10.1103/PhysRevB.99.094409}
  {\bibfield  {journal} {\bibinfo  {journal} {Phys. Rev. B}\ }\textbf {\bibinfo
  {volume} {99}},\ \bibinfo {pages} {094409} (\bibinfo {year}
  {2019})}\BibitemShut {NoStop}%
\bibitem [{\citenamefont {DuttaGupta}\ \emph {et~al.}(2017)\citenamefont
  {DuttaGupta}, \citenamefont {Fukami}, \citenamefont {Kuerbanjiang},
  \citenamefont {Sato}, \citenamefont {Matsukura}, \citenamefont {Lazarov},\
  and\ \citenamefont {Ohno}}]{DuttaGupta_AIPAdv_2017}%
  \BibitemOpen
  \bibfield  {author} {\bibinfo {author} {\bibfnamefont {S.}~\bibnamefont
  {DuttaGupta}}, \bibinfo {author} {\bibfnamefont {S.}~\bibnamefont {Fukami}},
  \bibinfo {author} {\bibfnamefont {B.}~\bibnamefont {Kuerbanjiang}}, \bibinfo
  {author} {\bibfnamefont {H.}~\bibnamefont {Sato}}, \bibinfo {author}
  {\bibfnamefont {F.}~\bibnamefont {Matsukura}}, \bibinfo {author}
  {\bibfnamefont {V.~K.}\ \bibnamefont {Lazarov}},\ and\ \bibinfo {author}
  {\bibfnamefont {H.}~\bibnamefont {Ohno}},\ }\href
  {https://doi.org/10.1063/1.4974889} {\bibfield  {journal} {\bibinfo
  {journal} {AIP Advances}\ }\textbf {\bibinfo {volume} {7}},\ \bibinfo {pages}
  {055918} (\bibinfo {year} {2017})}\BibitemShut {NoStop}%
\bibitem [{\citenamefont {Ma}\ \emph {et~al.}(2018)\citenamefont {Ma},
  \citenamefont {Yu}, \citenamefont {Tang}, \citenamefont {Li}, \citenamefont
  {He}, \citenamefont {Shi}, \citenamefont {Wang},\ and\ \citenamefont
  {Li}}]{Ma_PRL_2018}%
  \BibitemOpen
  \bibfield  {author} {\bibinfo {author} {\bibfnamefont {X.}~\bibnamefont
  {Ma}}, \bibinfo {author} {\bibfnamefont {G.}~\bibnamefont {Yu}}, \bibinfo
  {author} {\bibfnamefont {C.}~\bibnamefont {Tang}}, \bibinfo {author}
  {\bibfnamefont {X.}~\bibnamefont {Li}}, \bibinfo {author} {\bibfnamefont
  {C.}~\bibnamefont {He}}, \bibinfo {author} {\bibfnamefont {J.}~\bibnamefont
  {Shi}}, \bibinfo {author} {\bibfnamefont {K.~L.}\ \bibnamefont {Wang}},\ and\
  \bibinfo {author} {\bibfnamefont {X.}~\bibnamefont {Li}},\ }\href
  {https://doi.org/10.1103/PhysRevLett.120.157204} {\bibfield  {journal}
  {\bibinfo  {journal} {Phys. Rev. Lett.}\ }\textbf {\bibinfo {volume} {120}},\
  \bibinfo {pages} {157204} (\bibinfo {year} {2018})}\BibitemShut {NoStop}%
\bibitem [{\citenamefont {Zhang}\ \emph {et~al.}(2018)\citenamefont {Zhang},
  \citenamefont {Vernier}, \citenamefont {Zhao}, \citenamefont {Yu},
  \citenamefont {Vila}, \citenamefont {Zhang},\ and\ \citenamefont
  {Ravelosona}}]{Zhang_PRA_2018}%
  \BibitemOpen
  \bibfield  {author} {\bibinfo {author} {\bibfnamefont {X.}~\bibnamefont
  {Zhang}}, \bibinfo {author} {\bibfnamefont {N.}~\bibnamefont {Vernier}},
  \bibinfo {author} {\bibfnamefont {W.}~\bibnamefont {Zhao}}, \bibinfo {author}
  {\bibfnamefont {H.}~\bibnamefont {Yu}}, \bibinfo {author} {\bibfnamefont
  {L.}~\bibnamefont {Vila}}, \bibinfo {author} {\bibfnamefont {Y.}~\bibnamefont
  {Zhang}},\ and\ \bibinfo {author} {\bibfnamefont {D.}~\bibnamefont
  {Ravelosona}},\ }\bibfield  {title} {\bibinfo {title} {Direct observation of
  domain-wall surface tension by deflating or inflating a magnetic bubble},\
  }\href {https://doi.org/10.1103/PhysRevApplied.9.024032} {\bibfield
  {journal} {\bibinfo  {journal} {Phys. Rev. Applied}\ }\textbf {\bibinfo
  {volume} {9}},\ \bibinfo {pages} {024032} (\bibinfo {year}
  {2018})}\BibitemShut {NoStop}%
\bibitem [{\citenamefont {Ikeda}\ \emph {et~al.}(2010)\citenamefont {Ikeda},
  \citenamefont {Miura}, \citenamefont {Yamamoto}, \citenamefont {Mizunuma},
  \citenamefont {Gan}, \citenamefont {Endo}, \citenamefont {Kanai},
  \citenamefont {Hayakawa}, \citenamefont {Matsukura},\ and\ \citenamefont
  {Ohno}}]{Ikeda_NatMater_2010}%
  \BibitemOpen
  \bibfield  {author} {\bibinfo {author} {\bibfnamefont {S.}~\bibnamefont
  {Ikeda}}, \bibinfo {author} {\bibfnamefont {K.}~\bibnamefont {Miura}},
  \bibinfo {author} {\bibfnamefont {H.}~\bibnamefont {Yamamoto}}, \bibinfo
  {author} {\bibfnamefont {K.}~\bibnamefont {Mizunuma}}, \bibinfo {author}
  {\bibfnamefont {H.~D.}\ \bibnamefont {Gan}}, \bibinfo {author} {\bibfnamefont
  {M.}~\bibnamefont {Endo}}, \bibinfo {author} {\bibfnamefont {S.}~\bibnamefont
  {Kanai}}, \bibinfo {author} {\bibfnamefont {J.}~\bibnamefont {Hayakawa}},
  \bibinfo {author} {\bibfnamefont {F.}~\bibnamefont {Matsukura}},\ and\
  \bibinfo {author} {\bibfnamefont {H.}~\bibnamefont {Ohno}},\ }\href
  {https://doi.org/10.1038/nmat2804} {\bibfield  {journal} {\bibinfo  {journal}
  {Nature Materials}\ }\textbf {\bibinfo {volume} {9}},\ \bibinfo {pages} {721}
  (\bibinfo {year} {2010})}\BibitemShut {NoStop}%
\end{thebibliography}%

\end{document}